\documentclass[colorlinks=true, citecolor=blue, linkcolor=black]{hcig}

\usepackage{pdfpages}
\usepackage{xr-hyper}
\usepackage{hyperref}
\usepackage{amsmath} %
\usepackage{amsthm}  
\usepackage{amssymb}
\usepackage{color}
\usepackage{mathtools} 
\usepackage{multirow}
\usepackage{graphicx}
\graphicspath{{plots/}}
\usepackage{caption}
\usepackage[labelsep=colon]{caption} %
\usepackage{algorithm}
\usepackage{algpseudocode}
\usepackage{booktabs}
\usepackage{xcolor}
\usepackage{adjustbox}
\usepackage{subcaption}
\usepackage{tabularx}
\usepackage{svg}
\usepackage{wrapfig}

\usepackage{datetime}

\usepackage{url}
\usepackage{amsfonts}
\usepackage{nicefrac}
\usepackage{microtype}
\usepackage{bm}
\usepackage{mathrsfs}
\usepackage{cite}
\usepackage{enumitem}
\usepackage{hhline}
\usepackage{arydshln}
\usepackage{accents}
\usepackage{trimclip}
\usepackage{dsfont}
\usepackage{bbding}
\usepackage{setspace}
\usepackage{array, pifont}
\usepackage{makecell, multirow}
\usepackage{xcolor,colortbl}
\usepackage{afterpage}
\usepackage{pbox}
\allowdisplaybreaks

\definecolor{lightgreen}{rgb}{.9,1,.9}
\definecolor{hlcolor}{RGB}{220, 245, 220}
\definecolor{secondhlcolor}{RGB}{255, 238, 210}
\newcommand{\resultbest}[1]{\colorbox{hlcolor}{\strut \textbf{#1}}}
\newcommand{\resultsecond}[1]{\colorbox{secondhlcolor}{\strut #1}}

\newcommand{\cmark}{\ding{51}}%
\newcommand{\xmark}{\ding{55}}%
\newcommand{\halfcmark}{\cmark\textsuperscript{\kern-0.6em\raisebox{-0.48ex}{\xmark}}}

\newcolumntype{C}{>{\centering\arraybackslash}X}
\newcolumntype{W}{>{\centering\arraybackslash\hsize=1.2\hsize}X}
\newcolumntype{S}{>{\centering\arraybackslash\hsize=.9\hsize}X}

\newcolumntype{L}[1]{>{\raggedright\arraybackslash}p{#1}}
\newcolumntype{C}[1]{>{\centering\arraybackslash}p{#1}}
\newcolumntype{R}[1]{>{\raggedleft\arraybackslash}p{#1}}

\def\defn{\,\triangleq\,}
\def\argmin{\mathop{\mathsf{arg\,min}}} 

\def\lim{\mathop{\mathsf{lim}}} 
\def\min{\mathop{\mathsf{min}}}
\def\max{\mathop{\mathsf{max}}}

\def\log{\mathsf{log\,}}
\def\exp{\mathsf{exp}}

\def\bbm{{\bm{b}}}

\def\gbm{{\bm{g}}}

\def\qbm{{\bm{q}}}
\def\rbm{{\bm{r}}}

\def\tbm{{\bm{t}}}
\def\ubm{{\bm{u}}}

\def\xbm{{\bm{x}}}

\def\zbm{{\bm{z}}}
\def\zerobm{\bm{0}}

\def\Bbm{{\bm{B}}}

\def\Ibm{{\bm{I}}}

\def\C{\mathbb{C}}
\def\R{\mathbb{R}}

\def\Ebf{{\mathbf{E}}}

\def\Jbf{{\mathbf{J}}}

\def\fbf{{\mathbf{f}}}
\def\ybf{{\mathbf{y}}}

\def\Ccal{{\mathcal{C}}}

\def\Lcal{{\mathcal{L}}}

\def\Hcal{{\mathcal{H}}}

\def\Rcal{{\mathcal{R}}}

\DeclareMathAlphabet{\mathsfit}{T1}{\sfdefault}{\mddefault}{\sldefault}

\def\epsilon{\varepsilon}

\definecolor{limegreen}{rgb}{0.2, 0.8, 0.2}

\usepackage{xcolor}
\hypersetup{
    colorlinks,
    linkcolor={red!50!black},
    citecolor={blue},
    urlcolor={blue!80!black}
}

\title{ScoreField: Neural Inverse Scattering with Score-Based Generative Priors}

\author{Wenhan Guo, Yuan Gao, and Yu Sun\textsuperscript{\Letter}}
\address{Johns Hopkins University\\\smallskip
{\footnotesize \textsuperscript{\Letter}Corresponding author: ysun214@jh.edu}}

\headertitle{ScoreField}
\headerauthors{Guo et al.}

\begin{document}

\maketitle
\thispagestyle{firstpagestyle}

\begin{abstract}
Designing an effective electromagnetic inverse-scattering solver requires faithful enforcement of nonlinear full-wave physics together with an expressive prior on the unknown permittivity contrast. We propose \textit{ScoreField}, a neural inverse scattering framework that integrates coupled implicit neural representations (INRs) with a pretrained score-based generative prior. ScoreField employs two INRs to parameterize the permittivity contrast and the induced current fields, and jointly optimize them under the Lippmann–Schwinger equations. In addition to the implicit regularization by the INR architecture, the score model provides a learned prior gradient on the contrast, which is propagated to the contrast INR through the chain rule. This formulation enables ScoreField to effectively handle strong multiple scattering, where nonlinear wave interactions require accurate modeling of the coupled full-wave physics. We evaluate ScoreField on simulated weak- and strong-scattering benchmarks, the canonical Austria phantom, and experimental Fresnel measurements. We note that ScoreField significantly improves reconstruction fidelity and suppresses artifacts relative to classical full-wave methods and deep learning baselines, achieving an average PSNR improvement of $1.8 \, \mathrm{dB}$ over the best competing method on real Fresnel data.
\end{abstract}

\begin{figure*}
    \centering
    \includegraphics[width=1\linewidth]{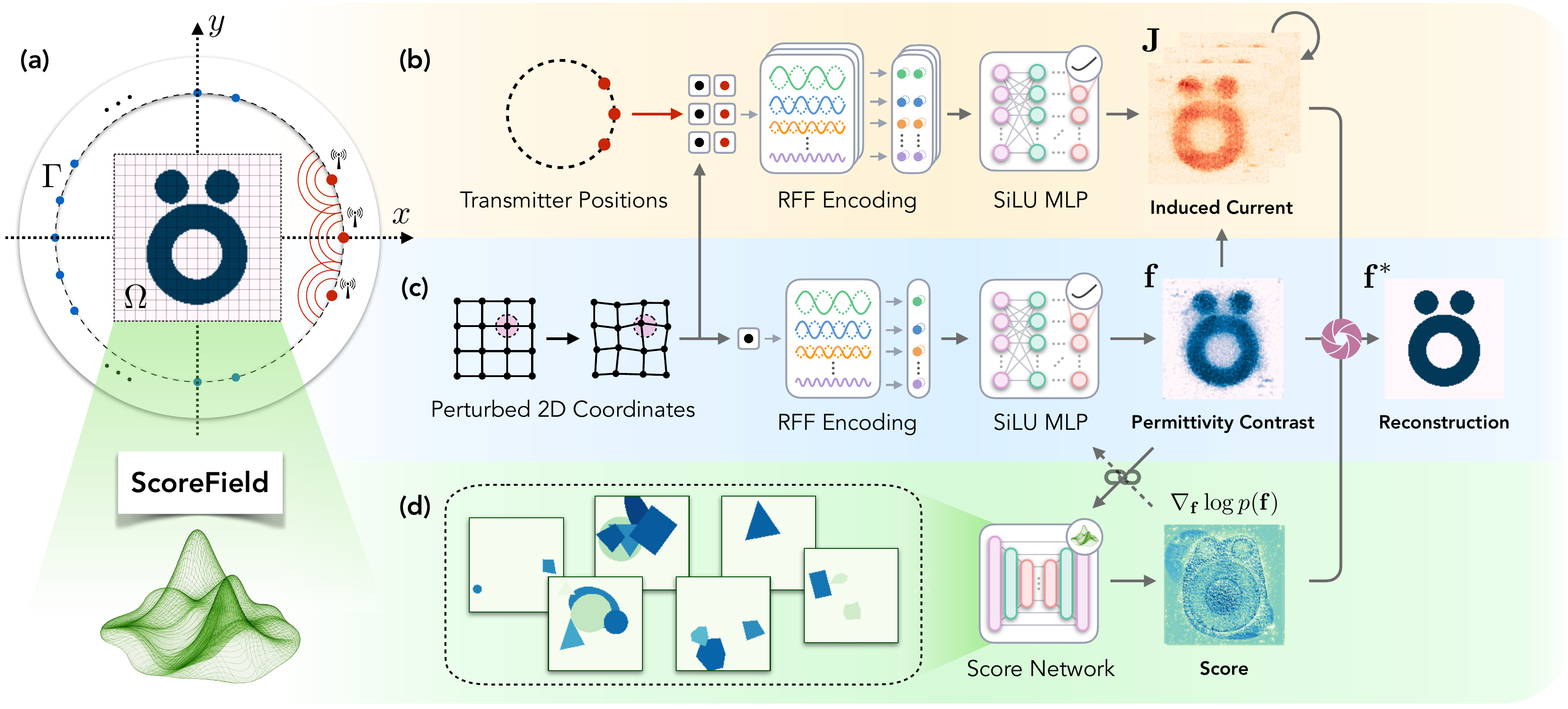}
    \caption{Conceptual overview of \emph{ScoreField}. \emph{(a)} Physical setup of inverse scattering considered in this work. The target in $\Omega$, described by contrast $\fbf(\rbm)$, is illuminated with an incident field $E^\mathsf{in}$ (\textit{red}). The resulting scattered field is measured at receiver locations on $\Gamma$ (\textit{blue}). \emph{(b)} Current INR: an illumination-dependent implicit neural representation (INR) predicts the induced current in the Lippmann--Schwinger equations~\eqref{eq:ls_state_continuous},~\eqref{eq:ls_data_continuous}. \emph{(c)} Contrast INR: another INR predicts the target contrast and is shared across illuminations. \emph{(d)} Score-based prior: a pretrained score model supplies a prior-gradient direction for the target contrast.}
    \label{fig:scorefield_teaser}
\end{figure*}

\section{Introduction}

Electromagnetic inverse scattering estimates a target's relative-permittivity contrast from measurements of its scattered electric field, with applications in optical diffraction tomography, radar imaging, and microwave biomedical tomography~\cite{Belkebir.Sentenac2003,Soto.etal2018,Maire.etal2009,Liu.etal2016a,Meaney.etal2000,zhu2022high}. In the weak-scattering regime, Born- or Rytov-type approximations replace the unknown internal field with a known reference field, yielding nearly linear computational imaging models~\cite{Wolf1969,Devaney1981,Kak.Slaney1988,Chen.Stamnes1998}. Many targets, however, violate this weak-scattering assumption. For high-contrast or electrically large targets, the wave undergoes multiple scattering inside the target before reaching the receivers, so the measured scattered field depends nonlinearly on both the target contrast and the unknown total field. This full-wave regime is governed by the Lippmann--Schwinger equation, and reconstruction must account for the self-consistent coupling between the target contrast and total field~\cite{Chew1999,Born.Wolf2003,Chew.Wang1990,Liu.etal2018,Ma.etal2018}.

Recovering the target contrast from scattered-field measurements defines a nonlinear inverse problem. Because finite and noisy measurements do not uniquely or stably determine the contrast, practical reconstruction methods solve a regularized optimization problem: a data-fidelity term enforces agreement between predicted and measured scattered fields, while a regularizer favors contrast images consistent with prior assumptions. Classical full-wave solvers~\cite{Soubies.etal2017,Pham.etal2018,Liu.etal2018,Ma.etal2018} express the wave physics through induced-current or contrast-source formulations and impose handcrafted penalties such as sparsity or total variation (TV). These penalties are interpretable and often effective for piecewise-smooth targets, but they encode generic local structure rather than target-class statistics. Consequently, they can oversmooth boundaries and textures and leave structured artifacts.

Learning-based inverse scattering methods provide richer priors, but they do not yet fully reconcile learned target-class statistics with full-wave physics. End-to-end networks such as ScaDec~\cite{Sun.etal2018} learn a direct map from simulated multiple-scattering data to the reconstructed target contrast and can be efficient at test time. Their learned inverse maps, however, are usually tied to the training distribution, acquisition geometry, and scattering regime. Implicit neural representation (INR) methods instead optimize coordinate-based models for each measurement. This instance-specific formulation is natural for inverse scattering because two INRs can represent the contrast and the induced current fields, thereby forming a joint learning problem coupled by the Lippmann--Schwinger equations~\cite{Mildenhall.etal2020,Tancik.etal2020,Luo2024ImagingInteriors}. Yet the regularization in these methods comes mainly from the neural architecture, coordinate encoding, and optimizer. It is not an explicit learned prior over plausible target contrasts.

Recently, score-based generative models have emerged as powerful and expressive image priors to capture complex distributions. The score function estimates the gradient of the log-density of an image distribution, and offers a principled way to introduce a learned prior without replacing the physical forward model for imaging inverse problems~\cite{Venkatakrishnan.etal2013,Romano.etal2017,Song.etal2019,Ho.etal2020,Song.etal2021score,Sun.etal2024,Wu.etal2024principled,alido2026whitened}. Motivated by this perspective, we propose \emph{ScoreField}, a neural inverse-scattering framework that integrates score-based generative priors into full-wave reconstruction. As illustrated in Fig.~\ref{fig:scorefield_teaser}, ScoreField optimizes two coupled INRs for the target contrast and the induced current. We utilize the Lippmann--Schwinger state and data equations~\eqref{eq:ls_state_continuous} and~\eqref{eq:ls_data_continuous} to enforce measurement consistency, while a pretrained score model supplies a learned prior direction on the contrast. In this way, the full-wave physics and the learned prior work together towards more accurate and expressive reconstruction.

Our main contributions are summarized as follows:
\begin{itemize}
    \item We formulate a neural inverse scattering framework in which a learned score prior guides the contrast reconstruction while the induced current remains constrained by full-wave physics.
    \item We instantiate this formulation as a practical test-time optimization algorithm with coupled contrast and current INRs, a full-wave loss, and score-based prior-gradient updates applied to the target contrast.
    \item We evaluate \textit{ScoreField} across strong and weak simulated scattering regimes and experimental Fresnel data, comparing against classical and learned solvers.
\end{itemize}
This article substantially extends our preliminary study~\cite{Gao.etal2025ScoreISP}. Instead of minimizing a loss on the norm of a pretrained score-model output, ScoreField adopts a more principled formulation to use the score itself as the prior-gradient direction for the target contrast. ScoreField also conducts systematic studies on strong- and weak-scattering simulations, real Fresnel data, out-of-distribution targets, broader baselines, and include robustness and calibration analyses. Across this broader evaluation, ScoreField improves reconstruction fidelity and reduces artifacts compared to the evaluated classical and learned baselines.

\section{Background}
\label{sec:background}

\subsection{Nonlinear Diffractive Imaging}

We consider two-dimensional nonlinear diffractive imaging in a homogeneous background medium. Let $\Omega \subset \R^2$ denote the bounded sample domain and let $\Gamma \subset \R^2$ denote the receiver orbit. The target is described by its permittivity distribution $\epsilon(\rbm)$, with known background permittivity $\epsilon_b$. We define the dimensionless target contrast
\begin{equation}
    f(\rbm) \defn \frac{\epsilon(\rbm)}{\epsilon_b} - 1, \qquad \rbm \in \Omega .
    \label{eq:contrast_definition}
\end{equation}
Throughout this paper we focus on the absorption-free setting, where the sampled target contrast is real-valued; the same physical model extends to complex-valued target contrast when absorption is included. For a monochromatic illumination, let $E^{\mathsf{in}}(\rbm)$, $E(\rbm)$, and $E^{\mathsf{sc}}(\rbm)$ denote the incident, total, and scattered electric fields, respectively. Under the scalar time-harmonic model, the total field inside the target domain satisfies the Lippmann--Schwinger equation~\cite{Chew1999,Born.Wolf2003,Chew.Wang1990,Liu.etal2018,Ma.etal2018,Yang.etal2020deep}
\begin{equation}
\label{eq:ls_state_continuous}
    E(\rbm)
    =
    E^{\mathsf{in}}(\rbm)
    +
    k_b^2 \int_{\Omega} g(\rbm-\rbm')
    \underbrace{f(\rbm')E(\rbm')}_{\defn\,J(\rbm')}
    \,d\rbm',
    \qquad \rbm\in\Omega,
\end{equation}
where $k_b = k_0\sqrt{\epsilon_b}$ is the background wavenumber, $k_0=2\pi/\lambda_0$ is the free-space wavenumber, and $J(\rbm)\defn f(\rbm)E(\rbm)$ is the induced current density. The two-dimensional outgoing Green's function is
\begin{equation}
    g(\rbm) \defn \frac{j}{4} H_0^{(1)}\left(k_b\|\rbm\|_{\ell_2}\right),
    \label{eq:greens_function_2d}
\end{equation}
where $H_0^{(1)}$ is the zero-order Hankel function of the first kind. The scattered field measured at receiver locations $\rbm\in\Gamma$ is the field radiated by the induced current,
\begin{equation}
\label{eq:ls_data_continuous}
    E^{\mathsf{sc}}(\rbm)
    =
    k_b^2 \int_{\Omega} g(\rbm-\rbm')J(\rbm')\,d\rbm'
    + e(\rbm),
    \qquad \rbm\in\Gamma,
\end{equation}
where $e$ denotes measurement noise.

Let $N$ be the number of spatial samples in $\Omega$, and let $M$ be the number of receiver samples on $\Gamma$. Discretizing~\eqref{eq:ls_state_continuous} and~\eqref{eq:ls_data_continuous} for a single illumination gives
\begin{subequations}
\label{eq:discrete_ls_single}
\begin{align}
    \Ebf &= \Ebf^{\mathsf{in}} + \mathbf G \, \Jbf,
    \label{eq:discrete_state_field} \\
    \Jbf &= \fbf \odot \Ebf,
    \label{eq:discrete_current} \\
    \ybf &= \mathbf S \, \Jbf + \mathbf e,
    \label{eq:discrete_data_field}
\end{align}
\end{subequations}
where $\fbf \in\R^N$ is the discretized contrast, $\Jbf \in \C^N$ is the discretized induced current, $\Ebf \in \C^N$, $\Ebf^{\mathsf{in}} \in \C^N$, and $\ybf \in \C^M$ represent $E$, $E^{\mathsf{in}}$, and $E^{\mathsf{sc}}$, respectively. The matrix $\mathbf G \in \C^{N \times N}$ evaluates the Green's function within $\Omega$, $\mathbf S \in \C^{M \times N}$ maps induced currents to the receiver orbit, $\odot$ denotes element-wise multiplication, and $\mathbf e \in \C^M$ is additive measurement noise. With multiple illuminations, the same contrast $\fbf$ is shared across illumination-dependent fields and currents; we use the single-illumination notation for clarity here.

Substituting the expression of the total field yields a nonlinear forward operator from contrast to data. For a given contrast $\fbf$, the induced current $\Jbf(\fbf)$ satisfies the fixed-point relation
\begin{equation}
    \Jbf(\fbf)
    =
    \fbf \odot \left(\Ebf^{\mathsf{in}} + \mathbf G\Jbf(\fbf)\right),
    \label{eq:current_fixed_point_background}
\end{equation}
and the measurement operator is
\begin{equation}
    \Hcal(\fbf) \defn \mathbf S\Jbf(\fbf),
    \qquad
    \ybf = \Hcal(\fbf) + \mathbf e .
    \label{eq:nonlinear_forward_operator_background}
\end{equation}
The inverse problem in~\eqref{eq:nonlinear_forward_operator_background} is nonlinear due to the dependence of $\Jbf(\fbf)$ on $\fbf$. Born- and Rytov-type approximations replace the unknown internal field by a fixed field, which is accurate only in weak-scattering regimes. For high-contrast or electrically large targets, multiple scattering makes this self-consistent coupling essential~\cite{Chew.Wang1990,Liu.etal2018,Sun.etal2018}.

\subsection{Bayesian Inference in Inverse Scattering}
\label{subsec:bayesian_inverse}

Equation~\eqref{eq:nonlinear_forward_operator_background} defines the inverse problem considered in this work: estimate the target contrast $\fbf$ from scattered-field measurements $\ybf$ using the known nonlinear operator $\Hcal$. Finite, noisy measurements generally do not determine $\fbf$ uniquely or stably, so prior information is needed to regularize the reconstruction. Bayesian inference combines the full-wave likelihood and the target-contrast prior through the posterior distribution~\cite{Tarantola2005,Molina.etal2001,Ribes.Schmitt2008}
\begin{equation}
    p(\fbf \,|\, \ybf)
    \propto
    p(\ybf\,|\,\fbf)\,p(\fbf).
    \label{eq:posterior_basic}
\end{equation}
When the receiver noise follows an \textit{i.i.d.} Gaussian distribution, and the prior has the form $p(\fbf)\propto \exp\{-\lambda\Rcal(\fbf)\}$, the maximum-a-posteriori (MAP) estimate is given by
\begin{equation}
\label{eq:map_estimator}
    \widehat{\fbf}_{\mathsf{MAP}}
    =
    \argmin_{\fbf \in \Ccal}
    \left\{
    \frac{1}{2}\left\|\ybf-\Hcal(\fbf)\right\|_{\ell_2}^2
    +
    \lambda\Rcal(\fbf)
    \right\}.
\end{equation}
Here $\Ccal$ may encode physical constraints such as nonnegativity or bounded contrast. Classical choices for $\Rcal$ include Tikhonov penalties, sparsity-promoting penalties, and total variation (TV), which remain common in inverse scattering and computational imaging~\cite{Beck.Teboulle2009a,Sung.Dasari2011,Liu.etal2018,Ma.etal2018}. Accordingly, TV-regularized full-wave solvers such as SEAGLE and CISOR can be interpreted as MAP estimators with edge-preserving handcrafted priors~\cite{Liu.etal2018,Ma.etal2018}. ScoreField retains the nonlinear Lippmann--Schwinger data model and replaces the explicit handcrafted penalty with a learned prior direction.

The posterior gradient makes the likelihood-prior separation explicit:
\begin{equation}
\label{eq:posterior_score_decomposition}
    \nabla_{\fbf}\log p(\fbf\,|\,\ybf)
    =
    \nabla_{\fbf}\log p(\ybf\,|\,\fbf)
    +
    \nabla_{\fbf}\log p(\fbf).
\end{equation}
The first term is determined by the nonlinear scattering model and measured scattered field; the second term is the score of the target contrast prior. This separation underlies plug-and-play priors~\cite{Venkatakrishnan.etal2013,Sreehari.etal2016} and regularization by denoising~\cite{Romano.etal2017,Reehorst.Schniter2019}, where physics-based data consistency is combined with a learned image model~\cite{Kamilov.etal2023}. ScoreField implements the two terms separately, where the gradients of the full-wave loss provide the likelihood gradient, and a pretrained score model supplies the prior gradient for the target contrast.

\subsection{Learning-Based Inverse Scattering}
\label{subsec:implicit_recon}

Most learning-based inverse-scattering methods either learn a direct reconstruction rule from training data or optimize a separate model for each measurement. Feed-forward and unrolled networks learn a map from measurements, or from a physics-informed transform of the measurements, to the desired contrast image. ScaDec~\cite{Sun.etal2018} backpropagates multiply scattered measurements into an image-domain representation and uses a U-Net decoder to recover the scattering potential. DeepNIS~\cite{Li.etal2019DeepNIS} uses a cascade of complex-valued residual CNN modules inspired by iterative nonlinear inverse scattering. SOM-Net~\cite{Liu.etal2022SOMNet} unrolls subspace-based optimization while embedding Lippmann--Schwinger consistency into the network design. These trained direct-reconstruction models can be efficient at test time, but the learned inverse map is tied to the training distribution, acquisition geometry, frequency range, and scattering regime. ScoreField instead optimizes each reconstruction against the measured full-wave model together with a learned contrast prior.

Instance-specific INR reconstruction optimizes a representation for each measurement rather than learning a single inverse map. For example, coordinate-based internal learning~\cite{Sun.etal2021a} showed that a coordinate MLP can be fit directly to a test target or its measurements without paired training data. More broadly, INRs parameterize images and physical fields as continuous coordinate-to-value maps, often using sinusoidal activations or Fourier-feature encodings to represent high-frequency structure~\cite{Mildenhall.etal2020,Sitzmann.etal2019,Tancik.etal2020,Liu.etal2022recovery}. This parameterization is well suited to inverse scattering because the target contrast is a spatial field and the induced current is an illumination-dependent wave field. Recent methods such as Imaging Interiors~\cite{Luo2024ImagingInteriors} represent both fields with INRs and optimize data and state losses derived from the Lippmann--Schwinger equations.
However, the architectural bias of an INR is not a distributional prior over scattering objects. ScoreField instead uses a learned score prior on the target contrast during instance-specific optimization, while the induced current remains constrained by the full-wave physics.

\subsection{Score-Based Generative Priors}
\label{subsec:score_prior_related}

Score-based generative models parameterize a family of Gaussian-smoothed image distributions by learning their score functions. Let $\xbm\sim p_{\mathsf{data}}$ denote a clean image and define its Gaussian perturbation as
\begin{equation}
    \xbm_\sigma = \xbm + \sigma\boldsymbol{\epsilon},
    \qquad
    \boldsymbol{\epsilon}\sim\mathcal N(\zerobm,\Ibm).
    \label{eq:gaussian_perturbation}
\end{equation}
Denoising score matching (DSM) trains a score model to approximate the score of the smoothed distribution $p_\sigma$~\cite{Vincent.etal2011,Song.etal2019,Ho.etal2020,Song.etal2021score}:
\begin{equation}
\label{eq:dsm_objective_background}
    s_\phi(\xbm_\sigma,\sigma)
    \approx
    \nabla_{\xbm_\sigma}\log p_\sigma(\xbm_\sigma).
\end{equation}
Equivalently, let $D_\phi(\zbm,\sigma)$ be a denoiser trained to predict a clean image from its Gaussian-corrupted observation. Tweedie's formula gives the relation~\cite{Efron2011,Laumont.etal2022}
\begin{equation}
    \nabla_{\zbm}\log p_\sigma(\zbm)
    \approx
    \frac{D_\phi(\zbm,\sigma)-\zbm}{\sigma^2}.
    \label{eq:tweedie_background}
\end{equation}
Thus, a denoiser defines a score model at noise scale $\sigma$. As $\sigma\to0$, $p_\sigma$ approaches $p_{\mathsf{data}}$; under standard smoothness conditions and wherever $p_{\mathsf{data}}$ is positive, $\nabla\log p_\sigma$ correspondingly approaches the clean-image score $\nabla\log p_{\mathsf{data}}$~\cite{Vincent.etal2011}.

In inverse problems, a score model can provide a learned prior gradient while leaving the measurement model explicit. Diffusion-based inverse solvers condition the reverse process on measurements or combine score-model updates with explicit data-consistency steps for image restoration and computational imaging~\cite{Kawar.etal2022denoising,Chung.etal2023diffusion,Sun.etal2024,Wu.etal2024principled}. Scores have also been used to regularize coordinate-based scene representations; DiffusioNeRF~\cite{Wynn2023DiffusioNeRF}, for example, backpropagates score-based prior updates from rendered RGBD patches to the color and density fields of a neural radiance field~\cite{Mildenhall.etal2020}. ScoreField generalizes this principle from ray-based rendering to full-wave inverse scattering, where the physical model is more complex and nonlinear due to multiple scattering. The next section describes how ScoreField combines the score-model prior update with gradients of the full-wave loss.

\section{Proposed Method: ScoreField}
\label{sec:method}

ScoreField reconstructs each measurement instance by optimizing two coupled INRs: a \emph{contrast INR} for the unknown target contrast and an illumination-dependent \emph{current INR} for the induced current. The Lippmann--Schwinger data and state equations define a full-wave loss for the two INRs, while a frozen score model provides a learned prior-gradient direction for the target contrast. We adopt the single-illumination notation of Sec.~\ref{sec:background}; for multiple transmitters, the contrast INR is shared, and the current INR, data loss, and state loss are evaluated for each transmitter and averaged. Consistent with the posterior-gradient decomposition in Sec.~\ref{subsec:bayesian_inverse}, gradients of the full-wave loss provide the likelihood-gradient term, and the score model propagates the prior gradient to the contrast INR.

\subsection{Coupled INR Backbone}
\label{subsec:implicit_scorefield}

ScoreField represents two distinct unknowns: the target contrast and the illumination-dependent induced current. For the INR inputs, let $\rbm_n=(x_n,y_n)\in[0,1]^2$ denote the normalized spatial coordinate of grid point $n$, and let $\tbm_\ell=(t_{x,\ell},t_{y,\ell})\in\mathcal T\subset[-1,1]^2$ denote the normalized position of transmitter $\ell$. The contrast INR $\widehat f_\theta:[0,1]^2\to\R$ represents the real-valued target contrast and is shared across illuminations. The current INR $\widehat J_\eta:[0,1]^2\times\mathcal T\to\C$ represents the complex induced current for each transmitter. This separation follows the induced-current formulation of the Lippmann--Schwinger equations and recent INR-based inverse-scattering solvers~\cite{Liu.etal2018,Ma.etal2018,Luo2024ImagingInteriors}.

Both INRs are coordinate MLPs with frozen random Fourier-feature (RFF) encodings. Plain coordinate MLPs are biased toward low-frequency functions, which can blur sharp target-contrast boundaries and underrepresent oscillatory current patterns~\cite{Rahaman.etal2019}. We adopt RFF encodings to expose the MLPs to a prescribed range of spatial frequencies~\cite{Tancik.etal2020,Mildenhall.etal2020,Sitzmann.etal2019}. Let $\ubm$ denote the input coordinate vector: $\ubm=\rbm$ for the contrast INR and $\ubm=[\rbm^\top,\tbm^\top]^\top$ for the current INR. We encode $\ubm$ as
\begin{equation}
    \gamma_{\Bbm}(\ubm)
    =
    \left[
        \sin(\Bbm\ubm+\bbm),\;
        \cos(\Bbm\ubm+\bbm)
    \right],
    \label{eq:scorefield_rff}
\end{equation}
where each row of $\Bbm$ is sampled independently from $\mathcal N(\zerobm,\sigma_B^2\Ibm)$ and every entry of the fixed phase vector $\bbm$ equals $\pi/4$. We sample $\Bbm$ once and freeze both $\Bbm$ and $\bbm$ throughout reconstruction. The resulting sinusoidal basis improves the representation of sharp target-contrast changes and oscillatory induced currents.

The contrast INR maps a spatial coordinate to a bounded target contrast,
\begin{equation}
    \widehat f_\theta(\rbm)
    =
    \frac{f_{\max}}{2}
    \left[
        1+\tanh\left(\Phi_\theta(\gamma_f(\rbm))\right)
    \right],
    \label{eq:scorefield_contrast_inr}
\end{equation}
where $\Phi_\theta$ is an MLP with SiLU activations and $f_{\max}$ is the maximum target contrast. We use SiLU because its smooth derivative supports stable differentiation through the INR and full-wave loss~\cite{Ramachandran.etal2017Swish,Elfwing.etal2018SiLU}.
The shifted $\tanh$ maps the unconstrained MLP output into $(0,f_{\max})$, thereby enforcing the nonnegative bounded-contrast model used in this paper.

The current INR maps a spatial coordinate $\rbm$ and transmitter coordinate $\tbm$ to a complex induced current. Let
\begin{equation}
    \Psi_\eta(\gamma_J([\rbm,\tbm]))
    =
    [\psi_{\eta,\mathrm{R}}(\rbm;\tbm),\psi_{\eta,\mathrm{I}}(\rbm;\tbm)]^\top
    \in\R^2 .
    \label{eq:scorefield_current_components}
\end{equation}
We define
\begin{equation}
    \widehat J_\eta(\rbm;\tbm)
    =
    s_J
    \left(
        \psi_{\eta,\mathrm{R}}(\rbm;\tbm)
        +
        j\psi_{\eta,\mathrm{I}}(\rbm;\tbm)
    \right),
    \label{eq:scorefield_current_inr}
\end{equation}
where $\Psi_\eta$ is a second SiLU MLP and $s_J$ is a fixed output scale for the current INR. The scale $s_J$ is part of the INR parameterization and is distinct from the state-loss weight $\lambda_s$ in~\eqref{eq:scorefield_physical_loss}. Evaluating both INRs on the grid gives the discrete variables used by the full-wave loss:
\begin{align}
    \fbf_\theta
    &=
    [\widehat f_\theta(\rbm_1),\ldots,
    \widehat f_\theta(\rbm_N)]^\top,
    \notag\\
    \Jbf_\eta(\tbm)
    &=
    [\widehat J_\eta(\rbm_1;\tbm),\ldots,
    \widehat J_\eta(\rbm_N;\tbm)]^\top .
    \label{eq:scorefield_grid_fields}
\end{align}
For notational brevity, we write $\Jbf_\eta$ for $\Jbf_\eta(\tbm)$ when a single illumination is under discussion.
During optimization, we query both INRs at perturbed spatial coordinates
\begin{equation}
    \widetilde{\rbm}_n=\rbm_n+\boldsymbol{\xi}_n,
    \qquad
    \boldsymbol{\xi}_n\sim\mathcal N(\zerobm,\sigma_r^2\Ibm),
    \label{eq:scorefield_coordinate_jitter}
\end{equation}
with a small $\sigma_r$. Coordinate jitter discourages grid memorization and grid-aligned artifacts by requiring the INRs to fit a neighborhood of each grid point rather than only the fixed pixel lattice~\cite{Tancik.etal2020,Luo2024ImagingInteriors}. The coupled INRs therefore parameterize $\fbf$ and $\Jbf$ as continuous fields while still producing the discrete vectors required by the Lippmann--Schwinger solver.

\begin{figure*}[t!]
    \centering
    \includegraphics[width=1\linewidth]{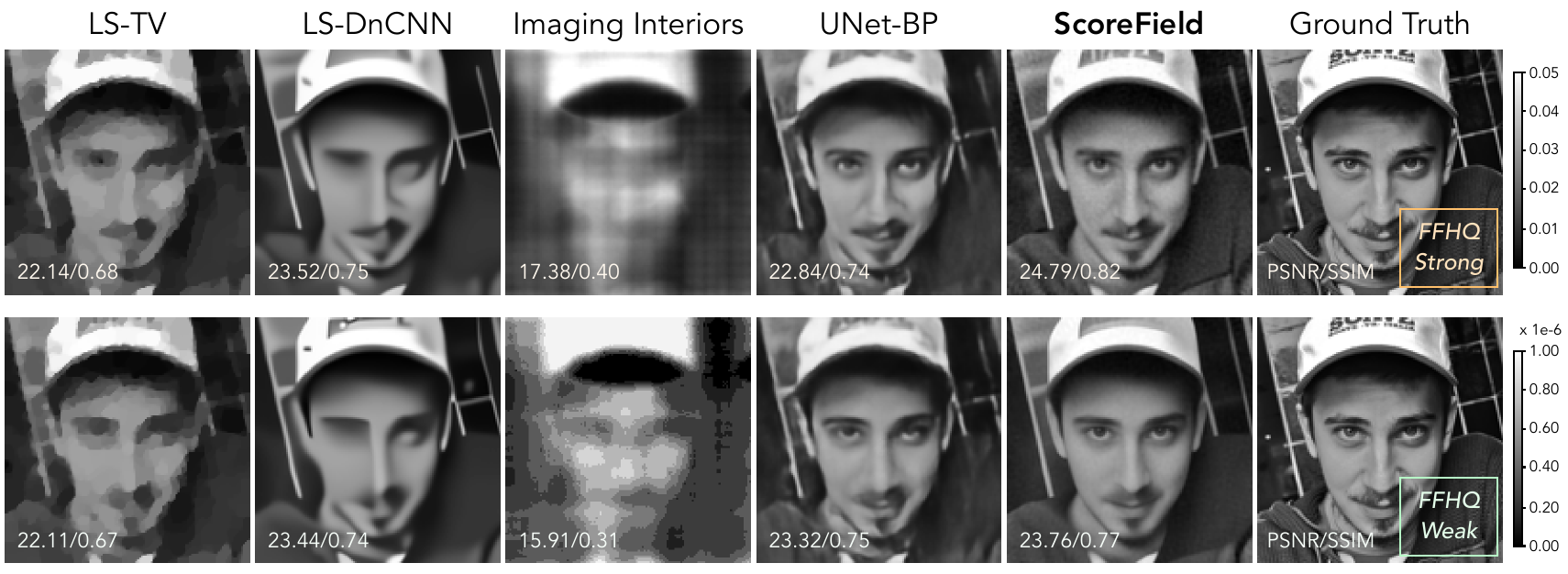}
    \caption{Visual comparison on simulated FFHQ inverse-scattering data under strong ($f_{\max}=0.05$) and weak ($f_{\max}=10^{-6}$) scattering. 
    Each image is labelled with its PSNR and SSIM values with respect to the groundtruth. 
    ScoreField improves facial detail and suppresses artifacts compared to other baselines. In particular, ScoreField better reconstructs the nose and mouth regions and preserves the rectilinear background patterns.}
    \label{fig:sim_ffhq}
\end{figure*}

\subsection{Physics-Constrained Optimization}
\label{subsec:physics_optimization}

The full-wave loss enforces the receiver equation and the induced-current state equation. Given $\fbf_\theta$ and $\Jbf_\eta$, ScoreField forms the predicted scattered-field measurements and the state-implied current as
\begin{equation}
    \widehat{\ybf}=\mathbf S\Jbf_\eta,
    \qquad
    \widehat{\Jbf}
    =
    \fbf_\theta\odot
    \left(\Ebf^{\mathsf{in}}+\mathbf G\Jbf_\eta\right).
    \label{eq:scorefield_forward_quantities}
\end{equation}
The normalized data and state losses are
\begin{subequations}
\label{eq:scorefield_physics_losses}
\begin{align}
    \Lcal_{\mathsf{data}}(\theta,\eta)
    &=
    \frac{
        M^{-1}\left\|\mathbf S\Jbf_\eta-\ybf\right\|_{\ell_2}^2
    }{
        M^{-1}\left\|\ybf\right\|_{\ell_2}^2+\epsilon_y
    },
    \label{eq:scorefield_data_loss}
    \\
    \Lcal_{\mathsf{state}}(\theta,\eta)
    &=
    \frac{
        N^{-1}\left\|
        \Jbf_\eta-\fbf_\theta\odot
        \left(\Ebf^{\mathsf{in}}+\mathbf G\Jbf_\eta\right)
        \right\|_{\ell_2}^2
    }{
        N^{-1}\left\|\fbf_\theta\odot\Ebf^{\mathsf{in}}\right\|_{\ell_2}^2+\epsilon_J
    } .
    \label{eq:scorefield_state_loss}
\end{align}
\end{subequations}
The factors $M^{-1}$ and $N^{-1}$ account for receiver count and grid size. The state loss uses the mean-squared magnitude of the Born current $\fbf_\theta\odot\Ebf^{\mathsf{in}}$ as its reference scale. We do not normalize by $\|\Jbf_\eta\|^2$ or $\|\widehat{\Jbf}\|^2$ because either choice would make the denominator depend on the current INR and would conflate current rescaling with minimization of the state residual. This normalization follows induced-current full-wave optimization, where residual magnitudes vary with scattering strength and acquisition geometry~\cite{Liu.etal2018,Ma.etal2018,Luo2024ImagingInteriors}. For masked receiver geometries, such as the experimental Fresnel data, $M$ counts the active measurements and $\Lcal_{\mathsf{data}}$ is evaluated only on those entries.

The full-wave loss combines measurement agreement and state consistency:
\begin{equation}
    \Lcal_{\mathsf{phys}}(\theta,\eta)
    =
    \lambda_y \, \Lcal_{\mathsf{data}}(\theta,\eta)
    +
    \lambda_s \, \Lcal_{\mathsf{state}}(\theta,\eta),
    \label{eq:scorefield_physical_loss}
\end{equation}
where $\lambda_y$ and $\lambda_s$ are fixed weights. The data term constrains the predicted scattered field at the receivers, and the state term enforces the induced-current fixed-point relation inside $\Omega$. The gradients of $\Lcal_{\mathsf{phys}}$ form the likelihood-side term in~\eqref{eq:scorefield_guided_gradients}.

An equivalent formulation of~\eqref{eq:discrete_ls_single} could let one INR predict $\fbf$ and another predict the internal scattered field $\Ebf^{\mathsf{sc}}_\Omega$, giving $\widehat{\ybf}_E=\mathbf S[\fbf\odot(\Ebf^{\mathsf{in}}+\Ebf^{\mathsf{sc}}_\Omega)]$. In this parameterization, both INRs enter the receiver prediction directly, and the data gradient for either INR depends on the current output of the other through the product $\fbf\odot\Ebf^{\mathsf{sc}}_\Omega$; more details can be seen in the supplementary material. Predicting $\Jbf$ instead follows induced-current and contrast-source optimization methods~\cite{Liu.etal2018,Ma.etal2018,Luo2024ImagingInteriors}: $\widehat\ybf=\mathbf S\Jbf_\eta$ is linear in the current INR and independent of the contrast INR, while the nonlinear coupling between target contrast and total field is confined to the state residual inside $\Omega$.

\subsection{Score-Based Contrast Prior}
\label{subsec:score_guidance}

The score model supplies learned target-contrast statistics as a prior-gradient direction. ScoreField normalizes the current target-contrast estimate to $[-1,1]$, denoted by $\tilde\fbf$, and evaluates a pretrained score model
\begin{equation}
    s_\phi(\tilde\fbf,\sigma)
    \approx
    \nabla_{\tilde\fbf}\log p_\sigma(\tilde\fbf),
    \label{eq:scorefield_score_model}
\end{equation}
where $p_\sigma$ is the target-contrast distribution smoothed by Gaussian noise of scale $\sigma$~\cite{Vincent.etal2011,Song.etal2019,Ho.etal2020,Song.etal2021score}. The score is an ascent direction for the smoothed log prior, while the full-wave measurement model remains explicit~\cite{Sun.etal2024,Wu.etal2024principled}.

The prior-gradient direction must be mapped from normalized image space to contrast-INR parameter space. Since the optimizer updates $\theta$ rather than pixels directly, ScoreField applies the adjoint Jacobian of $\tilde\fbf$:
\begin{equation}
    \Delta_\theta^{\mathsf{score}}(k)
    =
    \alpha_k
    \left(
        \frac{\partial \tilde\fbf}{\partial\theta}
    \right)^*
    s_\phi(\tilde\fbf,\sigma_k).
    \label{eq:scorefield_score_guidance}
\end{equation}
Here $(\cdot)^*$ denotes the adjoint Jacobian. In implementation, the pretrained score model is evaluated without gradient tracking, and only its output vector is propagated through the contrast-INR Jacobian.
The combined optimizer gradients are
\begin{subequations}
\label{eq:scorefield_guided_gradients}
\begin{align}
    \gbm_\theta
    &=
    \nabla_\theta\Lcal_{\mathsf{phys}}(\theta,\eta)
    -
    \Delta_\theta^{\mathsf{score}}(k),
    \\
    \gbm_\eta
    &=
    \nabla_\eta\Lcal_{\mathsf{phys}}(\theta,\eta).
\end{align}
\end{subequations}
Adam applies a descent step to $\gbm_\theta$, so subtracting $\Delta_\theta^{\mathsf{score}}$ from the supplied gradient produces ascent in the learned log prior. 

\begin{algorithm}[t]
\caption{ScoreField Reconstruction}
\label{alg:scorefield}
\begin{algorithmic}[1]
\Require $\ybf,\Ebf^{\mathsf{in}}$, operators $\mathbf G,\mathbf S$, grid $\{\rbm_n\}_{n=1}^N$
\Require frozen $s_\phi$, weights $\lambda_y,\lambda_s$, schedules $\alpha_k,\sigma_k$, physics-only warm-up $k_{\mathsf{warm}}$
\Ensure $\widehat{\fbf}$
\State Initialize $(\theta,\eta)$.
\For{$k=1,\ldots,K$}
    \State $\widetilde{\rbm}_n\gets\rbm_n+\boldsymbol{\xi}_n$.
    \State Query $\fbf_\theta$ and $\Jbf_\eta$.
    \State Compute $\widehat{\ybf}$ and $\widehat{\Jbf}$ by~\eqref{eq:scorefield_forward_quantities}.
    \State Compute $\Lcal_{\mathsf{phys}}(\theta,\eta)$.
    \State $(\gbm_\theta,\gbm_\eta)\gets\nabla_{(\theta,\eta)}\Lcal_{\mathsf{phys}}$.
    \If{$k>k_{\mathsf{warm}}$}
        \State Normalize $\fbf_\theta$ to $\tilde\fbf$ and set $\qbm\gets c(\sigma_k)s_\phi(\tilde\fbf,\sigma_k)$.
        \State $\gbm_\theta\gets
        \gbm_\theta-\alpha_k
        \left(\partial \tilde\fbf/\partial\theta\right)^*
        \qbm$.
    \EndIf
    \State Update $(\theta,\eta)$ with $\mathtt{Adam}$ optimizer.
\EndFor
\State \Return $\widehat{\fbf}=\fbf_\theta$ on $\{\rbm_n\}_{n=1}^N$.
\end{algorithmic}
\end{algorithm}
Algorithm~\ref{alg:scorefield} summarizes the reconstruction loop. Full-wave-loss gradients update both INRs, whereas the score-model prior gradient modifies only the contrast-INR update.

\begin{figure*}[t!]
    \centering
    \includegraphics[width=1\linewidth]{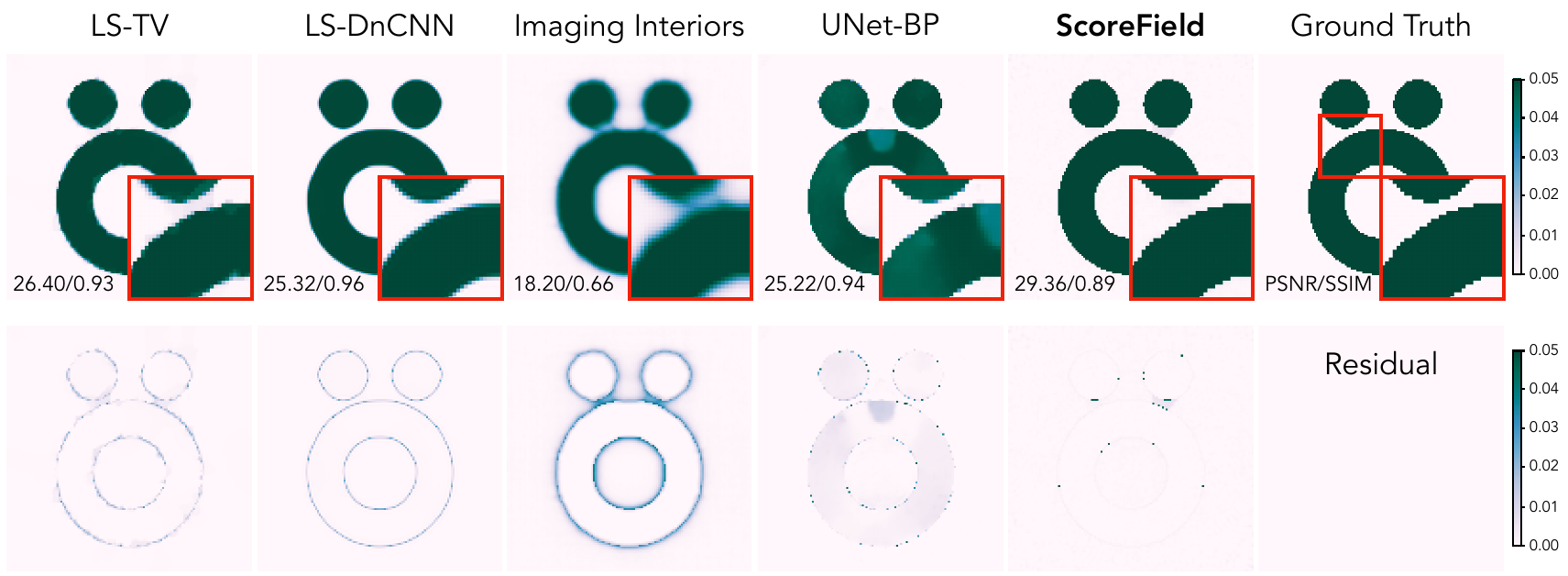}
    \caption{Visual comparison on the Austria profile with residual and zoomed boundary views. With a mismatched prior, ScoreField preserves the ring, disks, and interior void with small residuals concentrated near target boundaries. ScoreField has the least visible residual and best reconstructs both the sharp boundaries and the piecewise-constant structure.}
    \label{fig:sim_austria}
\end{figure*}

\subsection{Practical Design Choices}
\label{subsec:implementation}

The weights $\lambda_y$ and $\lambda_s$ set the relative importance of measurement agreement and state consistency. We linearly warm up the learning rate and $\lambda_s$. Because $\widehat{\ybf}=\mathbf S\Jbf_\eta$ does not depend on the contrast INR, starting $\lambda_s$ from zero lets the current INR first reduce the scattered-field data loss before the state term couples the two INRs. We pretrain the score model on target-contrast images normalized to $[-1,1]$ and keep it frozen during reconstruction. Let $k_{\mathsf{warm}}$ denote the number of physics-only iterations. For $k\leq k_{\mathsf{warm}}$, the score model is not evaluated, so neither INR receives a score-model prior gradient. For $k>k_{\mathsf{warm}}$, ScoreField computes the score-model prior gradient in~\eqref{eq:scorefield_score_guidance}. With $m=k-k_{\mathsf{warm}}-1$ denoting the zero-based prior-update index, the noise level and prior weight are
\begin{equation}
    \sigma_k = \max\left\{\sigma_0 \, \zeta_\sigma^m, \sigma_{\min}\right\},
    \quad
    \alpha_k = \max\left\{\alpha_0 \, \zeta_\alpha^m, \alpha_{\min}\right\},
\label{eq:annealing_schedule}
\end{equation}
where $0<\zeta_\sigma,\zeta_\alpha\leq1$ are decay factors. The schedules decrease the smoothing scale and prior weight until their respective lower bounds are reached. The numerical choices of INR size, RFF bandwidth, warm-up length, and optimization schedules are reported in the supplementary material.

\begin{table*}[t!]
\renewcommand{\arraystretch}{1.3}
\caption{Numerical results on simulated datasets. The best results are highlighted in \resultbest{green}, and the second best are highlighted in \resultsecond{orange}. ScoreField is the best or second best across all metrics on FFHQ strong, FFHQ weak, and Austria, except Austria SSIM; for this binary phantom, SSIM is less informative because small boundary shifts can dominate local-window scores.}
\centering
\label{tab:simulated_results}
\footnotesize
\setlength{\tabcolsep}{3pt}
\resizebox{\textwidth}{!}{%
\begin{tabular}{l cccc cccc cccc}
\toprule
\multirow{2}{*}{\textbf{Method}} &
\multicolumn{4}{c}{FFHQ \ $f_\mathrm{max} = 0.05$} &
\multicolumn{4}{c}{FFHQ \ $f_\mathrm{max} = 1\times10^{-6}$} &
\multicolumn{4}{c}{Austria \ $f_\mathrm{max} = 0.05$} \\
\cmidrule(lr){2-5}\cmidrule(lr){6-9}\cmidrule(lr){10-13}
&
PSNR$\uparrow$&
SSIM$\uparrow$&
MS-SSIM$\uparrow$&
LPIPS$\downarrow$&
PSNR$\uparrow$&
SSIM$\uparrow$&
MS-SSIM$\uparrow$&
LPIPS$\downarrow$&
PSNR$\uparrow$&
SSIM$\uparrow$&
MS-SSIM$\uparrow$&
LPIPS$\downarrow$\\
\midrule
BP&
6.51&0.009&0.100&0.869&
6.51&0.010&0.105&0.869&
5.68&0.570&0.140&0.756\\
FB-NN&
9.37&0.081&0.328&0.896&
14.54&0.271&0.811&0.773&
9.70&0.292&0.641&0.606\\
FB-TV&
11.96&0.290&0.398&0.473&
24.08&0.695&0.933&0.253&
10.29&0.384&0.703&0.315\\
LS-NN&
15.40&0.302&0.833&0.778&
14.54&0.271&0.811&0.773&
19.00&0.460&0.946&0.187\\
LS-TV&
24.38&0.710&0.939&0.242&
24.08&0.695&0.933&0.253&
\resultsecond{26.40}&0.926&\resultbest{0.998}&0.026\\
LS-DnCNN&
\resultsecond{25.26}&\resultsecond{0.760}&\resultsecond{0.953}&0.308&
24.89&\resultsecond{0.748}&\resultbest{0.948}&0.303&
25.32&\resultbest{0.956}&\resultbest{0.998}&0.077\\
UNet-BP&
24.74&0.750&0.941&\resultsecond{0.193}&
\resultsecond{25.01}&\resultbest{0.754}&\resultsecond{0.945}&\resultbest{0.185}&
25.22&\resultsecond{0.944}&\resultsecond{0.995}&\resultsecond{0.014}\\
Imaging Interiors&
18.77&0.497&0.800&0.637&
17.21&0.369&0.733&0.595&
18.20&0.662&0.970&0.296\\
ScoreField (ours)&
\resultbest{26.16}&\resultbest{0.786}&\resultbest{0.960}&\resultbest{0.181}&
\resultbest{25.14}&\resultbest{0.754}&\resultbest{0.948}&\resultsecond{0.225}&
\resultbest{29.36}&0.888&\resultbest{0.998}&\resultbest{0.006}\\
\bottomrule
\end{tabular}%
}
\end{table*}

\section{Experiments and Results}
\label{sec:experiments}

We evaluate ScoreField in five scenarios that combine three target classes, weak or strong scattering, and simulated or experimental acquisition geometries. The simulated scenarios comprise FFHQ~\cite{Karras.etal2019} target contrasts under strong scattering ($f_{\max}=5\times10^{-2}$) and weak scattering ($f_{\max}=10^{-6}$), together with the strong-scattering Austria profile; the experimental scenarios are the Fresnel FoamDielExt and FoamDielInt measurements. We report peak signal-to-noise ratio (PSNR), structural similarity (SSIM)~\cite{Wang.etal2004}, multiscale SSIM (MS-SSIM)~\cite{Wang.etal2003MS-SSIM}, and learned perceptual image patch similarity (LPIPS)~\cite{Zhang.etal2018LPIPS}; higher values are better for PSNR, SSIM, and MS-SSIM, whereas lower LPIPS is better. The baselines span backprojection (BP), first-Born methods with nonnegativity (FB-NN) and TV (FB-TV), full-wave methods with nonnegativity (LS-NN), TV (LS-TV), and a DnCNN denoiser (LS-DnCNN), an INR-based solver (Imaging Interiors), and an end-to-end supervised method (UNet-BP). Additional visual comparisons for all baselines are provided in the supplementary material.

\subsection{Score-Model Pretraining}
\label{subsec:score_training}

Before reconstruction, we pretrain one score model for the FFHQ prior and one for the polygon prior, then freeze both models. For the strong- and weak-scattering FFHQ experiments, we pretrain the FFHQ score model on $68{,}000$ target-contrast images and use $1{,}000$ additional images for validation; after normalization, the same model is used at both contrast scales. The score model is an attention U-Net based on the guided-diffusion architecture~\cite{Dhariwal.etal2021} and is trained by denoising score matching so that $s_\phi(\zbm,\sigma)$ approximates the score of the Gaussian-smoothed target-contrast distribution~\cite{Vincent.etal2011,Song.etal2019,Song.etal2021score}.
For the Austria and Fresnel experiments, we use the mismatched polygon prior described in Sec.~\ref{subsec:prior_mismatch}. We pretrain its score model on $68{,}000$ unpaired polygon target-contrast images and use $1{,}000$ images for validation, matching the FFHQ training protocol.

\begin{figure*}[h!]
    \centering
    \includegraphics[width=1\linewidth]{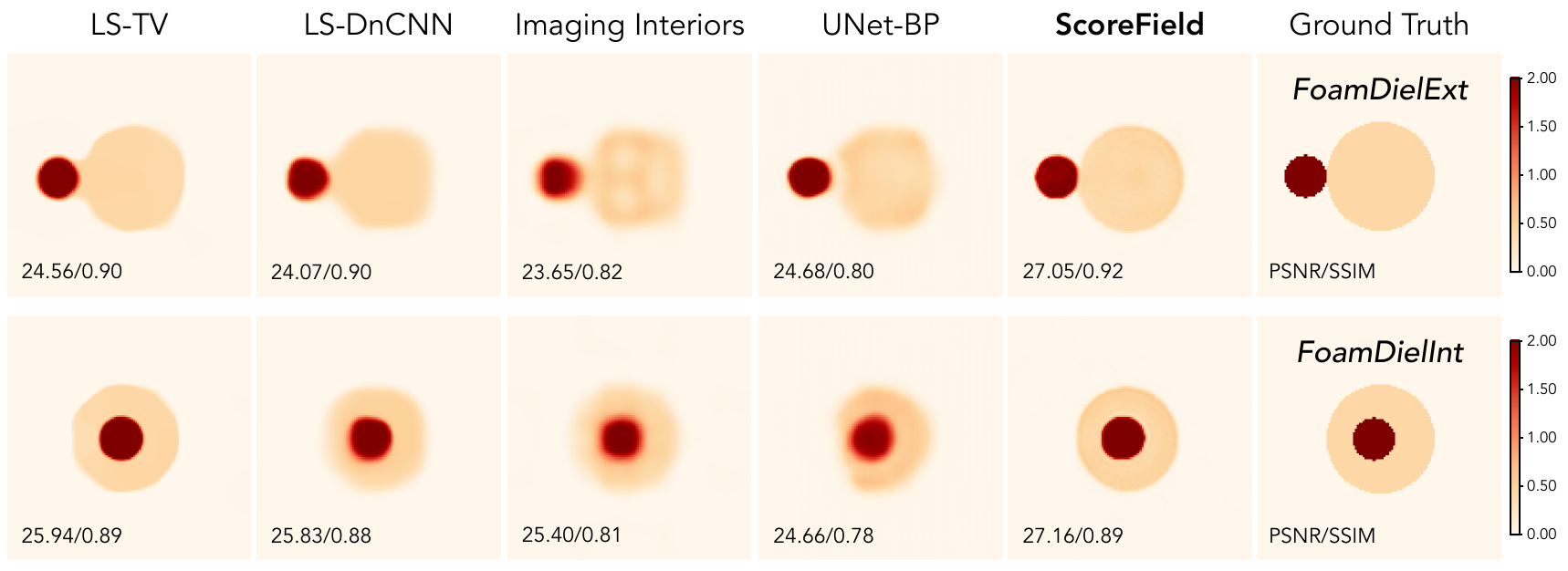}
    \caption{Visual comparison on experimental Fresnel \emph{FoamDielExtTM} and \emph{FoamDielIntTM} measurements. Each row shows one target geometry with the ground truth, classical Born and full-wave reconstructions, learned baselines, Imaging Interiors, and ScoreField. ScoreField recovers both the foam support and the high-contrast plastic inclusion with fewer visible artifacts, while LS-TV remains competitive on these piecewise-constant targets.}
    \label{fig:fresnel_results}
\end{figure*}

\subsection{Results on Simulated Datasets}

\paragraph{FFHQ Strong Scattering}
We first evaluate ScoreField on a strong-scattering FFHQ benchmark designed to test whether the learned contrast prior can help under nonlinear wave propagation. Each contrast image occupies an $18\,\mathrm{cm}\times18\,\mathrm{cm}$ field of view and is discretized on a $128\times128$ grid in a homogeneous background with $\epsilon_b=1$. The wavelength is $\lambda=0.84\,\mathrm{cm}$, with $40$ transmitters and $360$ receiver measurements on a circular array of radius $1.6\,\mathrm{m}$. We add measurement noise at $20\,\mathrm{dB}$ input SNR and scale the contrast to $0\leq f\leq f_{\max}$ with $f_{\max}=5\times10^{-2}$, producing a multiple-scattering regime in which Born-type linearization is inaccurate. All metrics are averaged over a separate set of $24$ held-out FFHQ images.

The paired Born and full-wave baselines in Fig.~\ref{fig:sim_ffhq} demonstrate the effectiveness of ScoreField. Full-wave solvers like LS-TV suppress artifacts by flattening facial texture and fine background structure. The learned baselines (LS-DnCNN, UNet-BP) produce more plausible faces, with LS-DnCNN restoring sharper local structure and UNet-BP giving a smooth supervised reconstruction, but UNet-BP requires paired training data for the same measurement setup and both methods still blur or distort fine features. Imaging Interiors captures part of the coarse support but leaves interior artifacts and does not reproduce the natural-image statistics of the face. ScoreField gives the cleanest reconstruction across facial structure, boundary sharpness, and artifact suppression, including the nose and mouth regions and the rectilinear background patterns.

Table~\ref{tab:simulated_results} confirms the same trend quantitatively. In the strong-scattering FFHQ scenario, ScoreField obtains the highest PSNR, SSIM, and MS-SSIM and the lowest LPIPS. Its PSNR is $0.90\,\mathrm{dB}$ above the second-best LS-DnCNN result and $1.42\,\mathrm{dB}$ above UNet-BP, while its LPIPS is $0.012$ below the next-best UNet-BP result. These gains across distortion, structural, multiscale, and perceptual metrics demonstrate the benefit of the score-model prior in full-wave reconstruction.

\paragraph{FFHQ Weak Scattering}
The weak-scattering FFHQ experiment uses the same geometry, physical setup, and score model as the strong-scattering experiment, but reduces the contrast scale to $f_{\max}=10^{-6}$. This setting is close to the linear scattering regime, where the first-Born approximation is accurate and the distinction between Born-based and full-wave data terms becomes small. This behavior is visible in Table~\ref{tab:simulated_results}: the first-Born nonnegative and TV baselines (FB-NN, FB-TV) match their full-wave counterparts (LS-NN, LS-TV) to the reported precision.

The weak-scattering visual comparison in the bottom two rows of Fig.~\ref{fig:sim_ffhq} shows a correspondingly narrower gap among the strongest methods. LS-TV recovers stable low-frequency facial structure but is still subject to significant piecewise-constant artifacts. Among the learned baselines, LS-DnCNN gives strong structural fidelity and UNet-BP produces smooth, perceptually plausible faces. ScoreField remains close to the ground truth while avoiding the most visible artifacts of the unregularized nonnegative baselines and the limited-representation artifacts seen in Imaging Interiors. The absolute PSNR and SSIM values are not higher than in the strong-scattering FFHQ experiment because the same $20\,\mathrm{dB}$ measurement-noise level becomes a dominant error source once the scattering nonlinearity is removed.

The numerical results in Table~\ref{tab:simulated_results} reflect this saturation effect. ScoreField obtains the best PSNR at $25.14\,\mathrm{dB}$, improving over UNet-BP by $0.13\,\mathrm{dB}$ and over LS-DnCNN by $0.25\,\mathrm{dB}$. It also matches the best SSIM value of $0.754$ and the best MS-SSIM value of $0.948$. UNet-BP has the lowest LPIPS in this weak-scattering case, with ScoreField second best. Thus, when the physics is nearly linear and several baselines are already strong, ScoreField still gives the best distortion and structural scores while remaining competitive perceptually.

\paragraph{Austria Profile}
We next test ScoreField on the standard Austria profile, a binary geometry consisting of a ring and two disks that is widely used as an inverse-scattering phantom. The physical acquisition is exactly the same as in the FFHQ strong-scattering experiment, with $f_{\max}=5\times10^{-2}$. ScoreField uses the mismatched polygon prior described in Sec.~\ref{subsec:prior_mismatch}.

The Austria comparison in Fig.~\ref{fig:sim_austria} emphasizes geometric fidelity rather than natural-image texture. Residuals are shown in the second row of Fig.~\ref{fig:sim_austria}. The zoomed $32\times32$ region highlights pixel-level boundary accuracy around the ring and interior void. The piecewise-constant Austria profile naturally favors TV regularization, and the TV-based baselines (LS-TV, Imaging Interiors) recover the global ring-and-disk layout. However, TV-regularized solvers can still thicken boundaries, shift contrast, or leave artifacts near the holes and disk edges. Learned baselines like LS-DnCNN and UNet-BP either produce over-smoothed boundaries or arbitrary variations in the contant contrast region. Despite using a mismatched polygon prior (see details in Sec.~\ref{subsec:prior_mismatch}, ScoreField gives the cleanest global shape and interior void, with residual errors mainly reduced to small boundary shifts visible in the zoomed view and residual map.

Table~\ref{tab:simulated_results} shows that ScoreField substantially outperforms the other methods on the Austria profile. It reaches $29.36\,\mathrm{dB}$ PSNR, a $2.96\,\mathrm{dB}$ gain over the second-best LS-TV result, obtains the lowest LPIPS value of $0.006$, and matches the best MS-SSIM value of $0.998$. Its SSIM is lower than that of LS-DnCNN and UNet-BP. However, small boundary displacements can reduce local-window SSIM even when the global support and perceptual error are better preserved. SSIM is therefore less informative than MS-SSIM and LPIPS in this case, where ScoreField preserves the global support with only a small number of shifted boundary pixels.

\begin{table*}[h!]
\renewcommand{\arraystretch}{1.3}
\caption{Numerical results on Fresnel datasets. The best results are highlighted in \resultbest{green}, and the second best are highlighted in \resultsecond{orange}. ScoreField achieves the best PSNR and MS-SSIM on both Fresnel targets and is the best or second best on every reported metric; LS-TV remains closest on SSIM and LPIPS for FoamDielInt because the target is nearly piecewise constant.}
\centering
\label{tab:fresnel_results}
\footnotesize
\setlength{\tabcolsep}{6.5pt}
\scalebox{1.107}{%
\begin{tabular}{l cccc cccc}
\toprule
\multirow{2}{*}{\textbf{Method}} &
\multicolumn{4}{c}{Fresnel FoamDielExt} &
\multicolumn{4}{c}{Fresnel FoamDielInt} \\
\cmidrule(lr){2-5}\cmidrule(lr){6-9}
&
PSNR$\uparrow$&
SSIM$\uparrow$&
MS-SSIM$\uparrow$&
LPIPS$\downarrow$&
PSNR$\uparrow$&
SSIM$\uparrow$&
MS-SSIM$\uparrow$&
LPIPS$\downarrow$\\
\midrule
BP&
15.05&0.744&0.530&0.578&
15.21&0.783&0.580&0.526\\
FB-NN&
15.16&0.603&0.570&0.501&
15.22&0.692&0.639&0.340\\
FB-TV&
15.33&0.521&0.654&0.444&
15.44&0.676&0.683&0.287\\
LS-NN&
22.77&0.721&0.911&0.351&
23.21&0.768&0.927&0.248\\
LS-TV&
24.56&\resultsecond{0.899}&\resultsecond{0.978}&\resultsecond{0.156}&
\resultsecond{25.94}&\resultbest{0.894}&\resultsecond{0.975}&\resultbest{0.082}\\
LS-DnCNN&
24.07&0.892&0.970&0.274&
25.83&0.884&0.972&0.240\\
UNet-BP&
\resultsecond{24.68}&0.799&0.966&0.259&
24.66&0.782&0.965&0.289\\
Imaging Interiors&
23.65&0.819&0.950&0.387&
25.39&0.808&0.966&0.319\\
ScoreField (ours)&
\resultbest{27.05}&\resultbest{0.917}&\resultbest{0.987}&\resultbest{0.102}&
\resultbest{27.15}&\resultsecond{0.892}&\resultbest{0.978}&\resultsecond{0.084}\\
\bottomrule
\end{tabular}%
}
\end{table*}

\subsection{Results on Experimental Datasets}

We evaluate ScoreField on the Institut Fresnel free-space experimental scattering database~\cite{Geffrin.etal2005,Geffrin.Sabouroux2009}. We use the TM-polarized FoamDielExt and FoamDielInt measurements, where a plastic cylinder is placed outside or inside a foam cylinder, respectively. The original database contains multiwavelength measurements from $2$ to $10\,\mathrm{GHz}$; in this work, we use only the $5\,\mathrm{GHz}$ measurements, for which the free-space wavelength is approximately $60\,\mathrm{mm}$. For FoamDielExtTM and FoamDielIntTM, the acquisition uses $8$ illuminations separated by $45^\circ$. For each illumination, the measured receivers lie on a partial circular arc: $241$ one-degree-spaced receiver locations are active, with receiver positions near the transmitter omitted. Our reconstruction operators use the processed $18\,\mathrm{cm}\times18\,\mathrm{cm}$ field of view on a $128\times128$ grid, the same receiver mask for all methods, and the contrast values $0.45$ for foam and $2$ for plastic.

Figure~\ref{fig:fresnel_results} compares reconstructions from the experimental Fresnel measurements; the first row shows FoamDielExtTM, and the second shows FoamDielIntTM. In this case, LS-TV performs well because Fresnel data is piecewise constant and strongly favor TV regularization. However, LS-TV still shows artifacts such as an erroneous smear between the two cylinders in FoamDielExt. The learned baselines (LS-DnCNN, UNet-BP) produce plausible supports but show residual contrast bias or boundary smoothing. Imaging Interiors captures the coarse layout, but its boundaries and interiors are less consistent across the two targets. ScoreField preserves both the foam cylinder and the high-contrast plastic inclusion with fewer visible artifacts.

Table~\ref{tab:fresnel_results} shows that ScoreField is the strongest method overall on the experimental data, while LS-TV remains a competitive baseline because of the piecewise-constant target geometry. On FoamDielExtTM, ScoreField is the best on all four metrics. Its PSNR is $2.37\,\mathrm{dB}$ higher than UNet-BP and $2.49\,\mathrm{dB}$ higher than LS-TV; it also improves LPIPS over LS-TV by $0.054$. On FoamDielIntTM, ScoreField outperforms all other baselines in PSNR and MS-SSIM, improving PSNR over LS-TV by $1.21\,\mathrm{dB}$ and over LS-DnCNN by $1.32\,\mathrm{dB}$. LS-TV and ScoreField obtain comparable SSIM and LPIPS on FoamDielIntTM, with only a $0.002$ margin for both metrics. Across the two Fresnel targets, ScoreField consistently exceeds all baselines in PSNR by more than $1\,\mathrm{dB}$, while remaining effectively tied with the strongest TV result on perceptual metrics, for which TV is favored by the target class.

\section{Robustness and Calibration Analysis}
\label{sec:robustness_calibration}

\subsection{Prior-Mismatch Robustness}
\label{subsec:prior_mismatch}

\begin{figure*}
    \centering
    \includegraphics[width=1\linewidth]{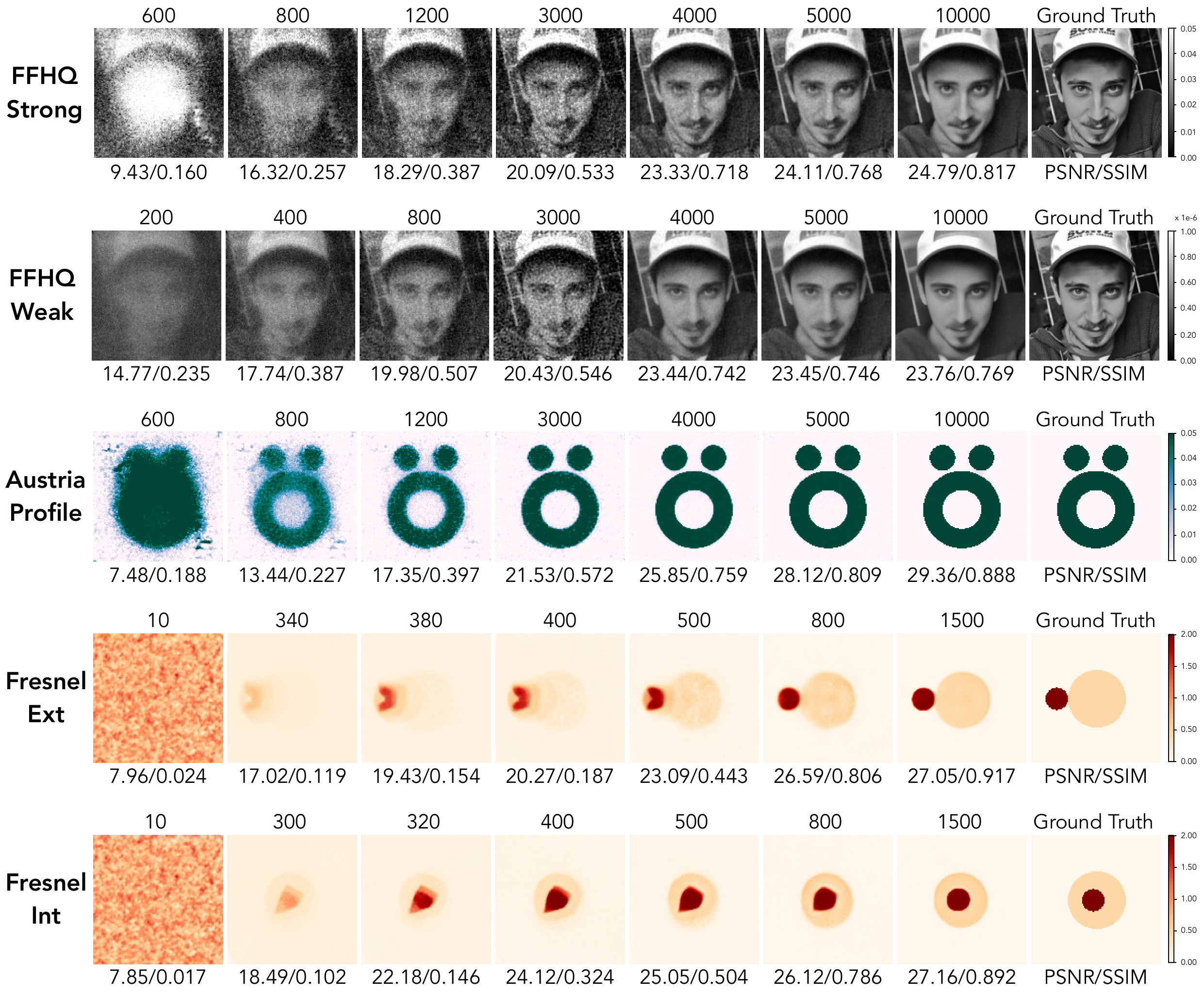}
    \caption{ScoreField reconstruction trajectories for representative simulated and experimental examples. We show the ScoreField iteration number on top of each intermediate reconstruction. For FFHQ strong, FFHQ weak, and Austria profile, the early iterations are dominated by the full-wave loss and begin to overfit noise before score guidance is injected at epoch $3000$, guiding the reconstruction toward sharper, higher-contrast images with fewer artifacts. For the Fresnel experiments, score guidance is injected early at epoch $10$, biasing the initial reconstruction toward the mismatched prior described in Sec.~\ref{subsec:prior_mismatch}, while the full-wave loss guides the reconstruction toward the true distribution. These trajectories show that the learned prior and measured full-wave data complement each other to improve reconstruction quality.}
    \label{fig:appendix_recon_process}
\end{figure*}

We test whether the learned prior remains useful when its training images differ from the reconstructed target. Supervised inverse maps can be sensitive to changes in target distribution, acquisition geometry, noise, or forward physics~\cite{McCann.etal2017,Lucas.etal2018,Han.etal2017,ZhangZ.etal2020}. For feed-forward inverse-scattering methods trained on paired simulations, such changes can require new measurement--target pairs~\cite{Sun.etal2018,Li.etal2019DeepNIS,Liu.etal2022SOMNet}. ScoreField instead pretrains its score model on unpaired target-contrast images and forms the full-wave loss from the measured scattered field during each reconstruction; the score model therefore does not encode a fixed inverse map for a particular acquisition geometry.

We test this setting with a deliberately mismatched polygon prior for the Austria and Fresnel experiments. Each $128\times128$ target-contrast training image contains $1$--$5$ randomly placed geometric primitives with intensities uniformly sampled from $(0,1]$. The primitive type is sampled from circles, ellipses, rotated rectangles, triangles, and irregular polygons with $3$--$8$ sides. The primitives may overlap, in which case the later primitive determines the intensity in the overlap. We pretrain the score model on $68{,}000$ polygon images and validate it on $1{,}000$ images; examples appear in the dashed box in the bottom row of Fig.~\ref{fig:scorefield_teaser}. This prior is intentionally mismatched: its images can be cluttered, overlapping, and variable in intensity, whereas the Austria and Fresnel targets contain regular circular structures with fixed contrasts and no random clutter.

The reconstruction trajectories in Fig.~\ref{fig:appendix_recon_process} show that the mismatched prior influences but does not determine the reconstruction. For Austria, the third row gradually settles into a sharp ring-and-disk geometry. For Fresnel, the fourth and fifth rows first form several overlapping polygon-like regions; continued minimization of the full-wave loss then moves them toward the measured two-cylinder geometry. This behavior suggests that the measured data can correct structures favored by a mismatched prior. By contrast, UNet-BP is trained separately for Austria and Fresnel on paired polygon simulations but remains more visibly affected by the mismatch in Figs.~\ref{fig:sim_austria} and~\ref{fig:fresnel_results}. The visual and numerical comparisons above show the same trend: ScoreField uses the unpaired polygon prior without being fixed to its training distribution.

\subsection{Model Calibration}

\begin{figure}[t!]
    \centering
    \includegraphics[width=0.65\linewidth]{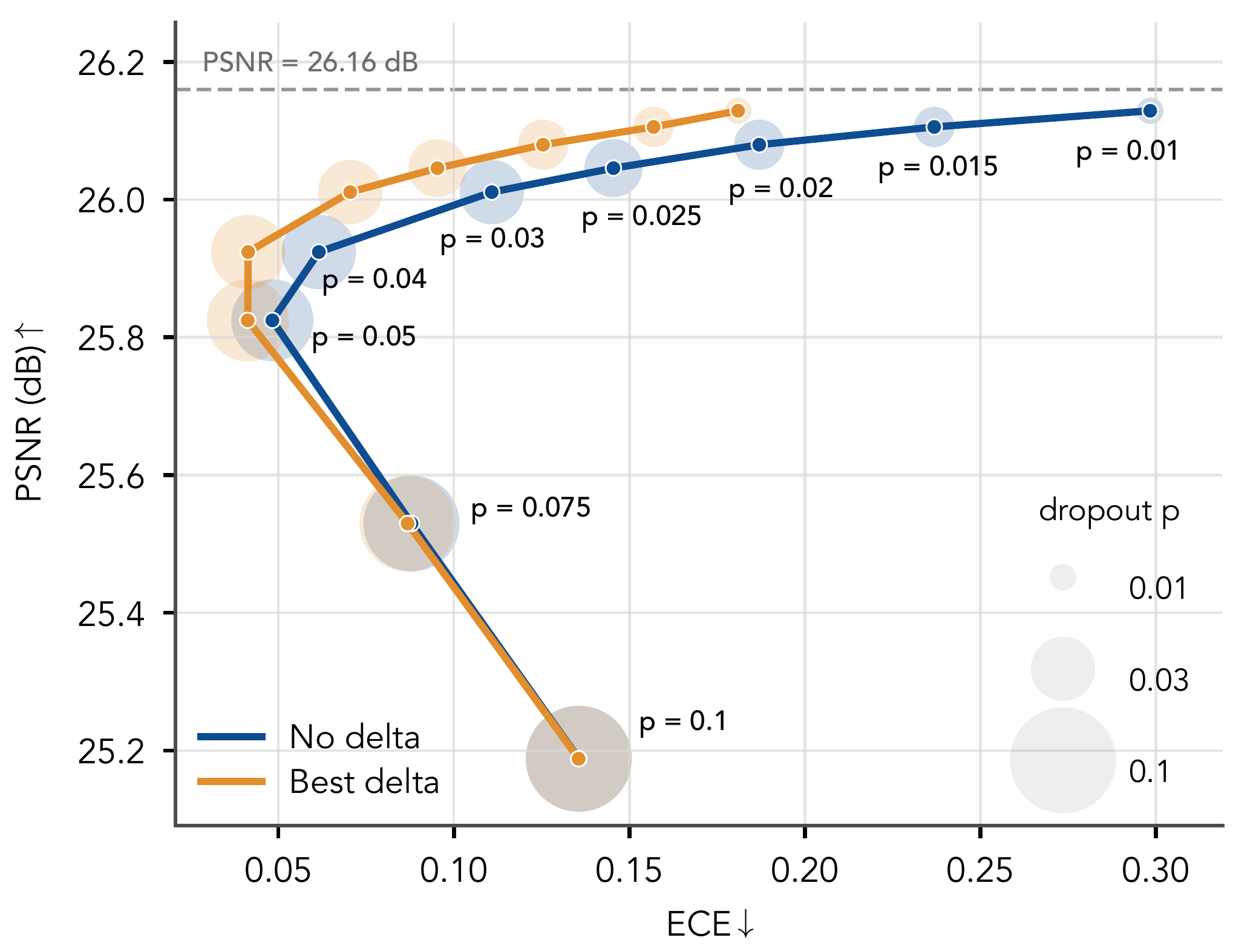}
    \caption{Monte Carlo dropout (MCD) calibration analysis on FFHQ strong scattering. Each point averages $24$ test reconstructions with $100$ dropout samples per image; marker size indicates the inference dropout probability $p$. Moderate dropout around $p=0.04$--$0.05$ gives the best calibration while keeping PSNR close to the deterministic reconstruction.}
    \label{fig:mcd_dropout}
\end{figure}

We next provide a complementary view of reconstruction reliability beyond image fidelity alone. In scientific reconstruction, and especially in inverse scattering, a reconstructed structure may support downstream physical interpretation, so the model's nominal confidence should be aligned with empirical correctness rather than only producing a visually plausible image. This perspective is closely related to the Bayesian formulation in Sec.~\ref{subsec:bayesian_inverse}, where uncertainty is part of the posterior description of the unknown contrast, and to prior work on uncertainty quantification for imaging inverse problems~\cite{Bardsley2012,Flath.etal2011,Repetti.etal2019}. We therefore include a preliminary calibration analysis for ScoreField as a diagnostic study to better understand the behavior of the method under stochastic perturbations.

We use Monte Carlo dropout (MCD) and expected calibration error (ECE) as lightweight tools for this analysis. MCD interprets dropout at test time as an approximate Bayesian ensemble, producing multiple predictions whose empirical spread can be used as an uncertainty proxy~\cite{Gal.2016a}. In this diagnostic, we enable dropout with probability $p$ in all layers of the contrast INR $\widehat f_\theta$, keep the current INR $\widehat J_\eta$ deterministic, and draw $100$ Monte Carlo samples for each of the $24$ strong-scattering FFHQ test images. Calibration is measured by comparing nominal predictive intervals with empirical coverage: for a calibrated reconstruction ensemble, a central interval with nominal level $q$ should contain approximately a $q$ fraction of ground-truth pixels. Following calibration practice in deep learning~\cite{Guo.etal2017Calibration,Gal.2016a} and recent INR uncertainty analysis~\cite{Vasconcelos.etal2023UncertaINR,SunHe.2021}, we compute ECE as the average absolute difference between nominal and observed coverage over levels from $0.05$ to $0.95$. We also report ECE-with-delta~\cite{Vasconcelos.etal2023UncertaINR}: we sweep a scalar interval expansion $\delta$ and record the lowest ECE, separating the reliability-curve shape from a global scale mismatch in the intervals. Both ECE measures are nonnegative and lower is better; zero indicates exact agreement between nominal and observed coverage at the evaluated levels.

Figure~\ref{fig:mcd_dropout} shows a clear accuracy--calibration tradeoff on FFHQ strong scattering. The deterministic ScoreField reconstruction in Table~\ref{tab:simulated_results} reaches $26.16\,\mathrm{dB}$ PSNR, while the MCD mean PSNR decreases as dropout increases, from $26.13\,\mathrm{dB}$ at $p=0.01$ to $25.19\,\mathrm{dB}$ at $p=0.10$. At the same time, calibration initially improves: the mean ECE drops from $0.298$ at $p=0.01$ to $0.048$ at $p=0.05$, and the best-delta ECE drops from $0.181$ to about $0.041$ around $p=0.04$--$0.05$. Larger dropout then degrades both reconstruction quality and calibration, with ECE rising to $0.136$ at $p=0.10$. Thus, for this FFHQ strong test, the most favorable operating region lies around $p=0.04$--$0.05$, where calibration improves substantially while PSNR remains within roughly $0.24$--$0.34\,\mathrm{dB}$ of the deterministic reconstruction. This experiment suggests that ScoreField has a tunable calibration--fidelity tradeoff, allowing it to be flexibly adapted to different application goals.
\section{Conclusion}
\label{sec:conclusion}

This paper introduced ScoreField, a score-regularized INR framework for nonlinear inverse scattering. The central idea is to represent the target contrast and illumination-dependent induced current with separate INRs, enforce the Lippmann--Schwinger data and state equations through a full-wave loss, and apply score-model prior-gradient updates to the contrast INR. Experiments on strong and weak FFHQ scattering, the Austria profile, and experimental Fresnel measurements show that this combination improves reconstruction fidelity and suppresses artifacts relative to classical and learned solvers. Future research can extend the reconstruction setting to three-dimensional inverse scattering with complex-valued target contrasts, or include broader acquisition geometries.

\section{Acknowledgement}
We acknowledge support from NSF Grant CCF-2542022.

\bibliographystyle{IEEEtran}
\bibliography{references}



\section*{Supplementary material}

# Supplementary Materials for
# *ScoreField: Neural Inverse Scattering with Score-Based Generative Priors*

## S1 Current and Scattered-Field Parameterizations

ScoreField predicts the induced current $\mathbf{J}$ rather than the internal scattered field $\mathbf{E}_\Omega^{\mathsf{sc}}$ as its auxiliary INR. This supplementary section records the algebra behind this choice: the two parameterizations are physically equivalent, zero initialization in either variable gives the Born approximation, and the current variable gives a simpler optimization objective.

### S1.1 Equivalent physical parameterizations

The induced current and the internal scattered field are equivalent variables for the Lippmann–Schwinger model. We suppress the illumination index and define

$$\mathbf{E}_\Omega^{\mathsf{sc}} \triangleq \mathbf{E} - \mathbf{E}^{\mathsf{in}}. \tag{S1}$$

The two variables satisfy

$$\mathbf{E}_\Omega^{\mathsf{sc}} = \mathbf{G}\mathbf{J}, \qquad \mathbf{J} = \mathbf{f} \odot (\mathbf{E}^{\mathsf{in}} + \mathbf{E}_\Omega^{\mathsf{sc}}), \qquad \mathbf{y} = \mathbf{S}\mathbf{J} + \mathbf{e}. \tag{S2}$$

Thus, one may predict $(\mathbf{f}, \mathbf{J})$ and recover $\mathbf{E}_\Omega^{\mathsf{sc}}$ by $\mathbf{G}\mathbf{J}$, or predict $(\mathbf{f}, \mathbf{E}_\Omega^{\mathsf{sc}})$ and recover $\mathbf{J}$ by $\mathbf{f} \odot (\mathbf{E}^{\mathsf{in}} + \mathbf{E}_\Omega^{\mathsf{sc}})$.

The two parameterizations have the same zero-residual solutions. In the current parameterization,

$$\widehat{\mathbf{y}}_J = \mathbf{S}\mathbf{J}, \qquad \boldsymbol{r}_J = \mathbf{J} - \mathbf{f} \odot (\mathbf{E}^{\mathsf{in}} + \mathbf{G}\mathbf{J}). \tag{S3}$$

In the scattered-field parameterization,

$$\widehat{\mathbf{y}}_E = \mathbf{S}\big[\mathbf{f} \odot (\mathbf{E}^{\mathsf{in}} + \mathbf{E}_\Omega^{\mathsf{sc}})\big], \qquad \boldsymbol{r}_E = \mathbf{E}_\Omega^{\mathsf{sc}} - \mathbf{G}\big[\mathbf{f} \odot (\mathbf{E}^{\mathsf{in}} + \mathbf{E}_\Omega^{\mathsf{sc}})\big]. \tag{S4}$$

If $\boldsymbol{r}_J = \mathbf{0}$, then $\mathbf{E}_\Omega^{\mathsf{sc}} = \mathbf{G}\mathbf{J}$ gives $\boldsymbol{r}_E = \mathbf{0}$ and the same receiver prediction. Conversely, if $\boldsymbol{r}_E = \mathbf{0}$, then $\mathbf{J} = \mathbf{f} \odot (\mathbf{E}^{\mathsf{in}} + \mathbf{E}_\Omega^{\mathsf{sc}})$ gives $\boldsymbol{r}_J = \mathbf{0}$ and the same receiver prediction. The choice of $\mathbf{J}$ or $\mathbf{E}_\Omega^{\mathsf{sc}}$ is therefore a choice of optimization coordinates, not a change of forward model.

### S1.2 Zero auxiliary fields and the Born approximation

Zero initialization in either auxiliary variable gives the first Born prediction for a fixed contrast. Starting the scattered-field fixed point from $\mathbf{E}_\Omega^{\mathsf{sc},0} = \mathbf{0}$ gives

$$\mathbf{E}_\Omega^{\mathsf{sc},1} = \mathbf{G}(\mathbf{f} \odot \mathbf{E}^{\mathsf{in}}), \qquad \mathbf{y}^1 = \mathbf{S}(\mathbf{f} \odot \mathbf{E}^{\mathsf{in}}). \tag{S5}$$

Starting the current fixed point from $\mathbf{J}^0 = \mathbf{0}$ gives

$$\mathbf{J}^1 = \mathbf{f} \odot \mathbf{E}^{\mathsf{in}}, \qquad \mathbf{E}_\Omega^{\mathsf{sc},1} = \mathbf{G}\mathbf{J}^1, \qquad \mathbf{y}^1 = \mathbf{S}\mathbf{J}^1. \tag{S6}$$

These are the same fields after substituting $\mathbf{J}^1 = \mathbf{f} \odot \mathbf{E}^{\mathsf{in}}$. They correspond to the first Born approximation, i.e., the leading term of the Born series before repeated internal scattering is added [1–3].

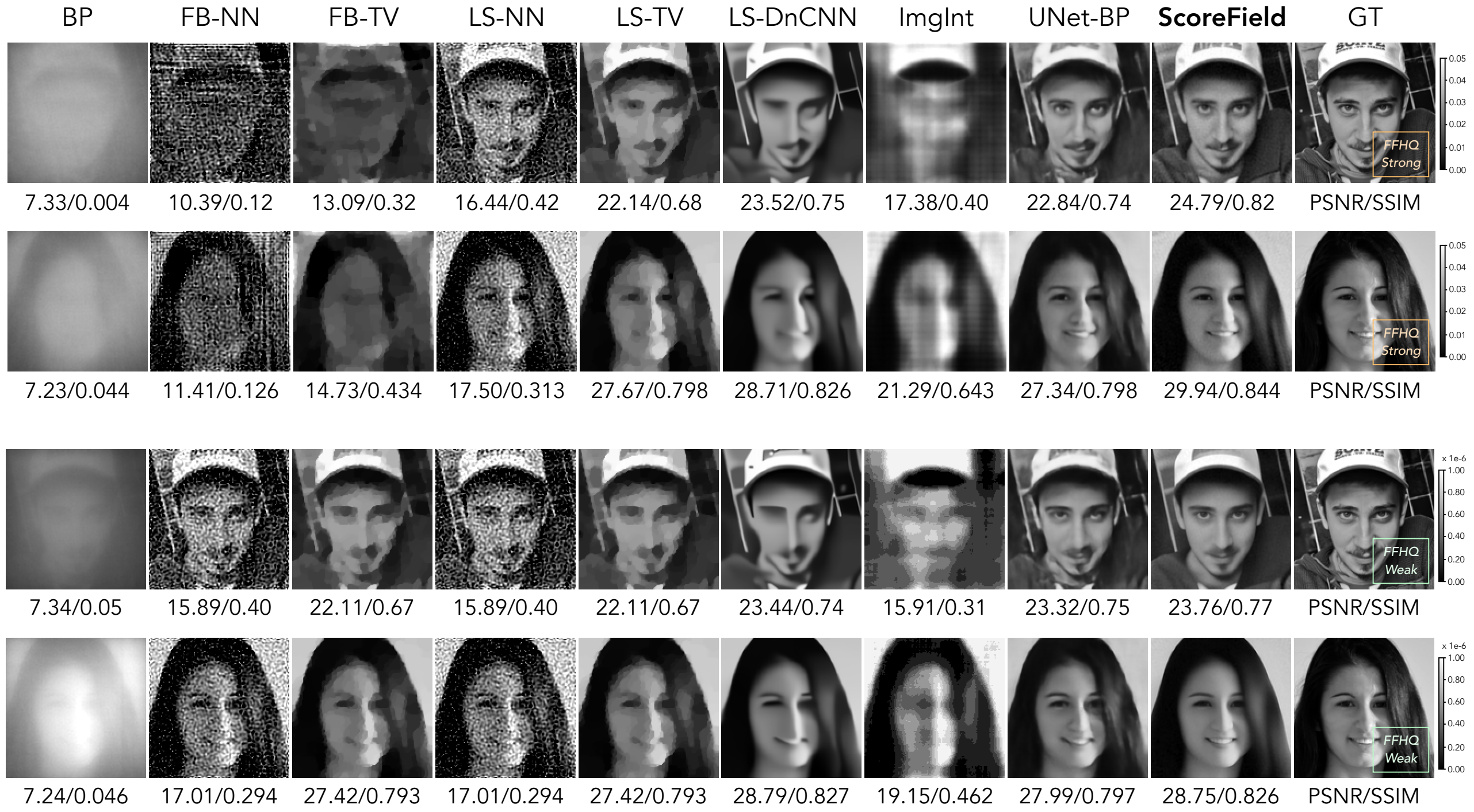


Figure S1: Full visual comparison for the simulated FFHQ and Austria experiments. The expanded views show that ScoreField consistently suppresses structured artifacts on FFHQ and yields the cleanest Austria support and residual map among the compared methods.

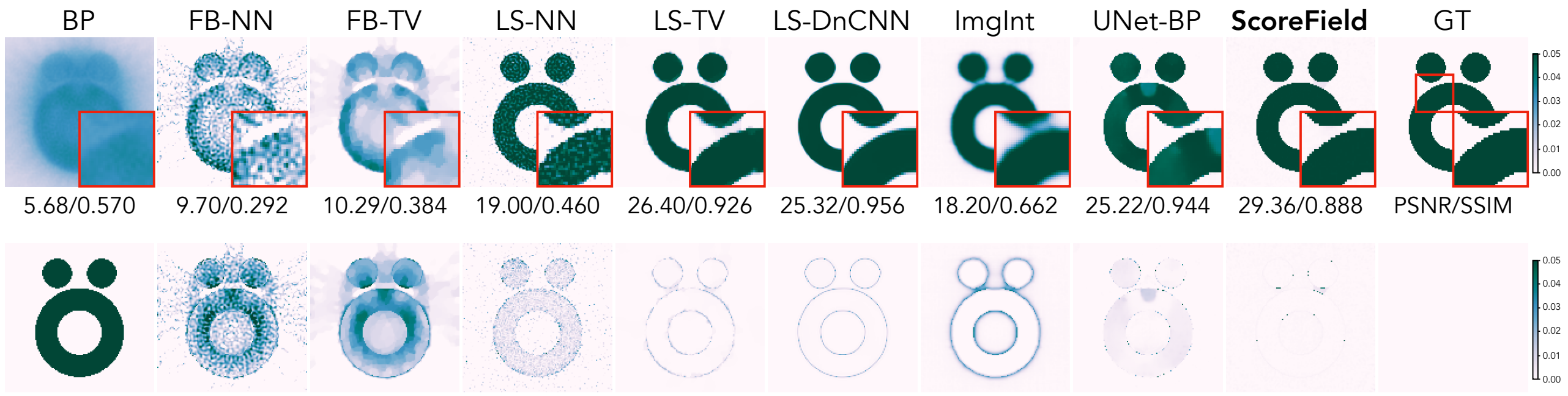


Figure S2: Full visual comparison for the simulated FFHQ and Austria experiments. The expanded views show that ScoreField consistently suppresses structured artifacts on FFHQ and yields the cleanest Austria support and residual map among the compared methods.

### S1.3 Why ScoreField predicts current

The current parameterization is cleaner because the data map is linear in the auxiliary field. For current,

$$\widehat{\mathbf{y}}_J = \mathbf{S}\mathbf{J}, \qquad \delta\widehat{\mathbf{y}}_J = \mathbf{S}\,\delta\mathbf{J}. \tag{S7}$$

Hence the data residual directly constrains the radiating source $\mathbf{J}$, while the contrast $\mathbf{f}$ enters through the state residual $\boldsymbol{r}_J$.

The scattered-field parameterization couples the contrast and auxiliary field inside the data map.

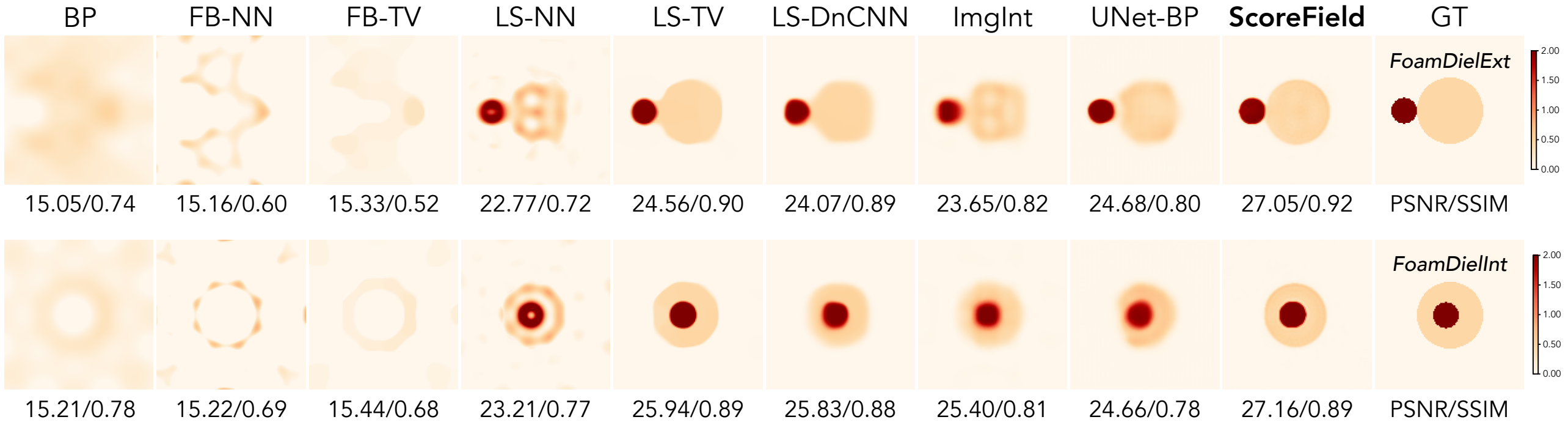


Figure S3: Full visual comparison for the Fresnel FoamDielExtTM and FoamDielIntTM experiments. ScoreField provides the most stable two-cylinder reconstructions across the exterior and interior plastic-inclusion settings, while TV remains strongest among the classical baselines.

Its first variation is

$$\delta\widehat{\mathbf{y}}_E = \mathbf{S}\big[\mathrm{diag}(\mathbf{E}^{\mathrm{in}} + \mathbf{E}^{\mathrm{sc}}_{\Omega})\delta\mathbf{f} + \mathrm{diag}(\mathbf{f})\delta\mathbf{E}^{\mathrm{sc}}_{\Omega}\big]. \tag{S8}$$

Thus, the data gradient for each INR is modulated by the current output of the other INR. This coupling is not erroneous, but it makes the receiver residual less direct for joint INR optimization.

The state residual also favors the current variable. The contrast variations are

$$\delta_{\mathbf{f}}\boldsymbol{r}_J = -\mathrm{diag}(\mathbf{E}^{\mathrm{in}} + \mathbf{G}\mathbf{J})\delta\mathbf{f}, \tag{S9}$$

$$\delta_{\mathbf{f}}\boldsymbol{r}_E = -\mathbf{G}\mathrm{diag}(\mathbf{E}^{\mathrm{in}} + \mathbf{E}^{\mathrm{sc}}_{\Omega})\delta\mathbf{f}. \tag{S10}$$

The current residual therefore has pointwise contrast dependence. In a least-squares state loss, the scattered-field residual backpropagates through the adjoint Green operator before reaching $\mathbf{f}$. Note that predicting $\mathbf{J}$ does not linearize inverse scattering; the state equation remains nonlinear in $(\mathbf{f}, \mathbf{J})$. It only separates the linear radiation map from the local coupling between target contrast and total field, which matches induced-current and contrast-source optimization methods [4–6].

## S2 ScoreField Hyperparameters

We use two INRs with SiLU activations and RFF to predict contrast and induced current, as described in the main paper. Unless otherwise noted, we use a contrast INR with depth 6, width 128, and 2048 RFF features, and a current INR with depth 6, width 256, and 4096 RFF features.

For FFHQ strong and Austria, we optimize ScoreField with RFF bandwidth 100, data weight $\lambda_y$=1.0, and state weight $\lambda_s$=0.3. We run Adam for 10,000 iterations with learning rate $5\times10^{-4}$ and linearly warm up the learning rate and $\lambda_s$ during the first 1000 iterations. The score model is not evaluated during the first 3000 iterations. Thereafter, we apply the score-model prior gradient with constant weight $\alpha_k$=$5\times10^{-7}$ and decay $\sigma_k$ from $\sigma_0$=80.0 to $\sigma_{\min}$=0.05 using $\zeta_\sigma$=0.999. For FFHQ weak scattering, we instead use learning rate $3\times10^{-3}$ and state weight $\lambda_s$=10.0, warm both quantities up for 200 iterations, and set $\alpha_k$=$2.0\times10^{-10}$.

The Fresnel runs use a smaller RFF configuration, with 512 contrast RFF features, 2048 current RFF features, and RFF bandwidth 30. We optimize each reconstruction for 1500 Adam iterations with learning rate $10^{-4}$, using a 50-iteration learning-rate warm-up. We set $\lambda_y$=1.0 and $\lambda_s$=2.0, with a 200-iteration warm-up for $\lambda_s$. The score model is not evaluated during the first 10 iterations. Thereafter,

the prior weight decays from $\alpha_0$=2.0×10$^{-4}$ to $\alpha_{\mathsf{min}}$=3.0×10$^{-5}$ using $\zeta_\alpha$=0.998, and the noise scale decays from $\sigma_0$=80.0 to $\sigma_{\mathsf{min}}$=0.5 using $\zeta_\sigma$=0.995.

## S3 Baseline Implementation Details

All baseline hyperparameters are fine-tuned with the same images used to tune ScoreField. For the FFHQ experiments, we tune FFHQ strong and FFHQ weak separately on three FFHQ images that are excluded from the 24-image test set. The Austria profile uses the same hyperparameters selected for FFHQ strong for every method, except for UNet-BP, which is retrained for the changed target distribution. For the Fresnel experiments, we jointly tune all methods on FoamDielExtTM and FoamDielIntTM.

**Back Projection (BP).** We compute

$$\mathbf{f}_{\mathsf{BP}} = \Re\left\{\sum_i \overline{\mathbf{E}_i^{\mathsf{in}}} \odot \mathbf{S}_i^H \mathbf{y}_i\right\}, \tag{S11}$$

where $i$ is the illumination index, the overline denotes complex conjugation, and $\mathbf{S}_i^H$ is the Hermitian adjoint of the discrete receiver Green's operator for that illumination. We keep the real part $\Re\{\cdot\}$ since contrast is real-valued.

**First Born + Non-Negativity (FB-NN).** We use the accelerated proximal-gradient update

$$\boldsymbol{v}_k = \boldsymbol{s}_k - \gamma \nabla D(\boldsymbol{s}_k), \tag{S12a}$$

$$\mathbf{f}_{k+1} = \mathsf{prox}_{\gamma R}(\boldsymbol{v}_k), \tag{S12b}$$

$$t_{k+1} = \frac{1+\sqrt{1+4t_k^2}}{2}, \tag{S12c}$$

$$\boldsymbol{s}_{k+1} = \mathbf{f}_{k+1} + \frac{t_k - 1}{t_{k+1}}(\mathbf{f}_{k+1} - \mathbf{f}_k). \tag{S12d}$$

where $\gamma$ is the stepsize. FB-NN minimizes the first-Born least-squares objective

$$D(\mathbf{f}) = \frac{1}{2}\|\mathbf{y} - \mathbf{S}(\mathbf{f} \odot \mathbf{E}^{\mathsf{in}})\|_{\ell_2}^2 \tag{S13}$$

with the nonnegative projection $R = \iota_{\mathbf{f}\geq 0}$. Thus, the proximal step in (S12b) is simply $\mathbf{f}_{k+1} = \mathsf{max}(\boldsymbol{v}_k, 0)$. We run FB-NN for 100 iterations.

**First Born + Total Variation (FB-TV).** FB-TV uses the same first-Born data term as FB-NN, but replaces the nonnegativity regularizer with TV and bound constraints, $R(\mathbf{f}) = \tau\,\mathsf{TV}(\mathbf{f}) + \iota_{\mathcal{C}}(\mathbf{f})$, where $\mathcal{C} = \{\mathbf{f} : 0 \leq f_n \leq \mathbf{f}_{\mathsf{max}},\ n = 1, \ldots, N\}$. The TV weight is parameterized as $\tau = m \cdot \frac{1}{2}\|\mathbf{y}\|_{\ell_2}$. For FFHQ strong and weak, we separately sweep

$$m \in \{10^{-7}, 3{\times}10^{-7}, 10^{-6}, 3{\times}10^{-6}, 10^{-5}, 3{\times}10^{-5}, 10^{-4}, 3{\times}10^{-4}, 10^{-3}\}$$

and select $m = 10^{-5}$ for both setups. For Fresnel, we use the same grid and select $m = 3{\times}10^{-4}$. We run FB-TV for 100 iterations.

**Lippmann–Schwinger + Non-Negativity (LS-NN).** LS-NN uses the full-wave data term [5]

$$D(\mathbf{f}) = \frac{1}{2}\|\mathbf{y} - \mathcal{H}(\mathbf{f})\|_{\ell_2}^2, \tag{S14}$$

where $\mathcal{H}$ evaluates the Lippmann–Schwinger forward model by solving for the total field. We use the same nonnegative projection used by FB-NN for the proximal step, and run LS-NN for 100 iterations.

**Lippmann–Schwinger + Total Variation (LS-TV).** LS-TV combines the full-wave data term with the TV proximal step [4,5]. The update is (S12) with $D$ in (S14) and $R(\mathbf{f}) = \tau\,\mathsf{TV}(\mathbf{f}) + \iota_{\mathcal{C}}(\mathbf{f})$. We sweep over the same TV grid $m$, and select $m = 10^{-5}$ for FFHQ strong and weak, and $m = 3{\times}10^{-4}$ for the jointly tuned Fresnel pair. We run LS-TV for 100 iterations.

**Lippmann–Schwinger + Denoiser (LS-DnCNN).** To the best of our knowledge, there is currently no standard plug-and-play inverse-scattering baseline that combines a learned denoiser with the full-wave Lippmann–Schwinger solver [7]. We therefore implement LS-DnCNN by replacing the proximal step in (S12b) with a pretrained DnCNN denoiser [8–10] within the same full-wave data term used by LS-NN and LS-TV. We normalize the contrast image range to $[0, 1]$ before denoising and rescale the denoised image back to the physical contrast range. We sweep over the grid $\gamma \in \{2, 5, 10, 20, 40, 80, 150, 300, 500, 800, 1000\}$ and denoiser noise level $\sigma \in \{1, 2, 3, 4, 5, 7, 8, 10, 12, 15\}$, selecting $\gamma = 500$ and $\sigma = 4$ for FFHQ strong and weak, and $\gamma = 40$ and $\sigma = 10$ for the Fresnel pair. We run LS-DnCNN for 100 iterations.

**Imaging Interiors.** We adapt the official Imaging Interiors method [6] to our physical setup. Importantly, the original implementation explicitly forms the dense domain Green's matrix $\Phi_{\mathsf{mat}} \in \mathbb{C}^{N\times N}$ and applies the internal field interaction as $\Phi_{\mathsf{mat}}\mathbf{J}$. This is reasonable under a $64 \times 64$ grid discretization, but becomes computationally infeasible at $128 \times 128$ since $\Phi_{\mathsf{mat}}$ would contain $(128^2)^2$ entries. We replace the dense multiplication with FFT convolution using the same domain Green's kernel. We validate our implementation on a $64{\times}64$ Austria profile using the original physical setup and reproduce the published Imaging Interiors behavior. We then tune the *learning rate* $\in \{5.0 \times 10^{-4}, 8.0 \times 10^{-4}, 1.0 \times 10^{-3}, 2.0 \times 10^{-3}, 5.0 \times 10^{-3}\}$, *current scale* $\in 1.0 \times \{10^{-9}, 10^{-8}, 10^{-7}, 10^{-6}, 10^{-5}, 10^{-4}, 10^{-3}\}$, *TV weight* $\in \{0.0, 1.0 \times 10^{-5}, 1.0 \times 10^{-4}, 1.0 \times 10^{-3}\}$, and positional encoding *bandwidth* $\in \{10, 15, 20\}$ on the same validation images used for the other baselines. We set *sample_perturb*$= -1$ to match the original relative coordinate perturbation to each physical field of view.

**UNet + Back Projection (UNet-BP).** UNet-BP follows the ScaDec pipeline [11], where a fixed incident-field backprojection produces a two-channel real/imaginary input, and a U-Net decoder predicts the contrast. We reimplement ScaDec in PyTorch and preserve its network structure. Because the supervised map is tied to the physical setup, we train four separate models for FFHQ strong, FFHQ weak, Austria, and Fresnel. The two FFHQ models are trained on 68,000 FFHQ backprojected images generated with the corresponding scattering strength; the Austria and Fresnel models are trained on 68,000 backprojected mismatched polygon images described in the main paper's prior-mismatch robustness analysis and generated with the corresponding physical setup. We use 1,000 validation images and train each model for 1000 epochs. For each training run, we use a batch size of 24 and tune over the coarse grid *learning rate* $\in \{2.0 \times 10^{-4}, 5.0 \times 10^{-4}, 1.0 \times 10^{-3}\}$, *dropout* $\in \{0.2, 0.25, 0.3\}$ and select the best checkpoint with early stopping.



\end{document}